\documentclass[12pt]{article}

\usepackage{epsfig}
\usepackage{amssymb}

\newcommand{\sect}[1]{\setcounter{equation}{0}\section{#1}}

\def\be{\begin{equation}}
\def\ee{\end{equation}}
\def\ba{\begin{eqnarray}}
\def\ea{\end{eqnarray}}

\title{{\bf Kaluza-Klein Branes}}
\author{J.F.Sparks\thanks{email: J.F.Sparks@damtp.cam.ac.uk}
\\ \\ DAMTP \\ Centre for Mathematical Sciences \\ University of
Cambridge \\ Wilberforce Road, Cambridge CB3 0WA, UK. 
\\ \\ Preprint DAMTP-2001-42}

\date{21 May 2001}

\begin{document}

\maketitle

\begin{abstract}

We examine Kaluza-Klein branes in detail. Specifically, we show that codimension four
submanifolds that are stationary under a semi-free circle action may
be interpreted as branes or antibranes in the Kaluza-Klein reduced
space that are magnetically charged under the Kaluza-Klein field
strength. We derive the equation in cohomology that is satisfied by such a brane using an
explicit construction of the Thom class of the normal bundle of the
brane worldvolume in the reduced space. This may be applied to both
the D6-brane of Type IIA String Theory, and also to various
recent constructions of magnetic branes immersed in fluxbrane backgrounds. We then go on to study
the special case of monopole-antimonopole production in a five-dimensional Kaluza-Klein
theory, illustrating our arguments with various concrete examples.

\end{abstract}

\sect{Introduction}

In \cite{horowitz}, various examples of magnetically charged
strings or $p$-branes were constructed, at the level of the
appropriate low-energy effective theory, where the gauge field derives from Kaluza-Klein reduction on a circle. More recently, various papers
have given further examples of this construction \cite{refs}, where
spherical, or more generally, tubular, branes are immersed (in
the physical rather than strict mathematical sense) in a background magnetic
fluxbrane; that is, the brane solution approaches the fluxbrane
solution asymptotically. The fluxbrane is essentially just a
generalisation of the Melvin Universe \cite{gibbons} and provides the
magnetic force required to prevent the brane from collapsing due to
its own tension. Such solutions
are of course typically unstable. 

In this paper, we focus on the case of Kaluza-Klein branes. By
definition, these are branes that are magnetically charged under a $U(1)$ Kaluza-Klein field
strength. In \cite{horowitz}, various examples of Kaluza-Klein branes were
constructed by considering a circle action on some manifold $X$, and
then Kaluza-Klein reducing on the circle direction; the branes are
identified with the stationary points of the circle action. This
construction produces both the Type IIA D6-brane, at the level of
supergravity, and also various examples of magnetic $p$-branes
immersed in fluxbrane backgrounds. The aim of the
present paper is to make the relation between stationary points of
circle actions and Kaluza-Klein branes more precise. In general, it is not
obvious how such stationary point sets may be interpreted as branes in
the reduced space, and in particular, why the corresponding branes may be charged under the
Kaluza-Klein field strength. The above facts were deduced in
\cite{horowitz}, and the barrage of recent papers \cite{refs}, by examining
specific examples, rather than giving a general argument. It is also
worth noting that in all these examples, and indeed in general, the dilaton diverges as one
approaches the brane. Hence, physically, the space is decompactifying
near the brane,
and one should therefore work in the higher dimensional spacetime in a
neighbourhood of the brane. However, in order to examine properties of the brane from the
point of view of the base spacetime, one needs to interpret the brane
as an object that is intrinsic to the base. We fill this gap in the literature. Assuming that the higher dimensional spacetime $X$ takes
the form $X=\mathbb{R}\times{M}$ where $\mathbb{R}$ is the time
direction and $M$ is the spatial manifold on which we dimensionally
reduce, we show that codimension four (with respect to $M$) submanifolds that are stationary under a semi-free circle action may
be interpreted as branes or antibranes in the reduced space that are
magnetically charged with respect to the Kaluza-Klein field strength,
$G_2$. Such a brane acts as a source for $G_2$

\be
[dG_2]=[\delta(W)]\label{brane}\ee

This holds\footnote{For an antibrane there is an extra
minus sign. We have normalised $G_2$ such that
the brane charge is $1$. More generally, one has
$[dG_2]=\pm{Q}[\delta(W)]$ for a brane (antibrane) of charge $Q$
($-Q$).} as an equation in the cohomology group $H^3(B)$ of the
spatial base $B$,
where $\delta(W)$ is a closed three-form that is
Poincar$\acute{\mathrm{e}}$ dual to the brane worldvolume $W$ in $B$, and has
support on $W$. We derive this equation from the Kaluza-Klein
perspective using an explicit construction of the Thom class of the
normal bundle of $W$ in $B$. The present paper therefore both formalises
and generalises previous work, placing the examples of
\cite{horowitz}, together with the examples contained in more recent
papers, in a general setting. Note that the reduced spacetime is
actually of the
form $\mathbb{R}\times{B}$. The time direction is topologically trivial, and in
particular will not enter into our topological considerations. We
therefore simply neglect the factor of $\mathbb{R}$ in most of the
paper, dealing either with the Riemannian manifold $M$, which is a
spacelike hypersurface in $X$, or the Kaluza-Klein reduction $B$ of $M$.

We then go on to use the above ideas in the context of
monopole-antimonopole production in a five-dimensional Kaluza-Klein
theory. In dimension four, Kaluza-Klein monopoles are $0$-branes which
are charged under the Kaluza-Klein $U(1)$ gauge field. The details of
monopole-antimonopole production are constrained by the topology of
the nucleation surface. In particular, the fact that the total number
of monopoles and antimonopoles must be equal (charge conservation),
and that the total number of defects produced is given by the Euler
characteristic of the nucleation surface, may be derived using various
$G$-index theorems. Using a result of Fintushel on the classification
of circle actions on simply-connected four-manifolds, together with
some standard cobordism results, we are able to classify completely
the possible topologies of the nucleation surface, assuming the latter
is simply connected. Finally, we give several examples which describe
the nucleation of Kaluza-Klein monopoles and antimonopoles either in
the presence of a positive cosmological constant, or a thin domain wall.

The plan of the paper is as follows. In Section 2, we motivate the
present paper by reminding the reader of some of the examples of
Kaluza-Klein branes contained in reference \cite{horowitz}. In Section
3, we analyse circle actions in detail. This section is somewhat
technical, and is included only to correct some of the misconceptions
in the current literature. In Section 4, we derive equation
(\ref{brane}) from two points of view. Firstly, we assume that we have
a codimension three brane $W$ living on a $d$-dimensional manifold $B$
that couples magnetically to $G_2$ via a Wess-Zumino coupling 

\be
(-1)^D\int_{\mathbb{R}\times{W}}C_{D-4}\label{coupling}\ee

in the 'string frame' action, where $C_{D-4}$ is the potential for the
dual field strength of $G_2$ and we
also define $\mathrm{dim}(X)=D$ and $\mathrm{dim}(M)=d+1$, so that $d=D-2$. The equation (\ref{brane}) is then derived by
a constrained variation of the total action. This is more or less
standard. The second point of view is to regard $W$
as arising, in a way described more precisely later, from a
codimension four (again, with respect to $M$) stationary point set of a semi-free
circle action of the higher dimensional space $M$. The Kaluza-Klein
2-form $G_2$ is not defined on the whole base $B$. We first show that
there exists an extension of $G_2$ on the whole space $B$ such that equation
(\ref{brane}) is satisfied; this is related to a specific construction of the Thom
class of the normal bundle of $W$ in $B$. We then show that this
extension is unique, as far as cohomology is concerned. We also give a detailed
account of brane charge, illustrating the discussion using some of the
examples in Section 2.

As mentioned, in Section 5 we study the case $D=5$ in detail. Kaluza-Klein branes
are simply monopoles in this case, and we use various theorems on
circle actions in order to deduce the qualitative details of
monopole-antimonopole pair production. Finally, in Section 6 we give
several explicit examples, illustrating the ideas of the previous
section. Our conclusions are contained in Section 7.

\sect{Motivation}

In this section, we briefly remind the reader of some of the examples of
Kaluza-Klein branes constructed in \cite{horowitz}. These will serve
both as motivation for the present paper, and also as illustrations of
some of the more abstract topological ideas we shall encounter later. 

\subsection{The Kaluza-Klein monopole}

The Kaluza-Klein monopole is a solution to the canonical five-dimensional
Kaluza-Klein theory. The monopole itself is a $0$-brane that is magnetically
charged under the $U(1)$ Kaluza-Klein gauge field. Specifically, the
Ricci-flat five-manifold $X$ is given by a metric product
$X=\mathbb{R}\times{M}$ where $\mathbb{R}$ is the time direction, and
$M$ is the Euclidean (anti-)self-dual Taub-NUT metric. Thus

\be
ds^2=-dt^2+ds^2_{\mathrm{Taub-NUT}}\label{monopole}\ee

where 

\be
ds^2_{\mathrm{Taub-NUT}}=\left(\frac{r+a}{r-a}\right)dr^2+4a^2\left(\frac{r-a}{r+a}\right)(d\psi+\cos\theta{d}\phi)^2+(r^2-a^2)(d\theta^2+\sin^2\theta{d}\phi^2)\label{taubnut}\ee

The Taub-NUT metric is hyperK$\ddot{\mathrm{a}}$hler, with holonomy $SU(2)$, and is therefore
Ricci-flat. The radial coordinate $r$ takes the range
$a\leq{r}<\infty$, and $(\psi,\theta,\phi)$ are Euler angles on
$S^3$. The manifold is topologically $\mathbb{R}^4$ and asymptotically
flat. 

The monopole solution (\ref{monopole}) admits a circle
isometry\footnote{We shall use the terms circle action and $U(1)$
action interchangeably.} generated by the Killing vector field
$\partial/\partial\psi$. This has a one-dimensional stationary point
set given by $\{r=a, -\infty<t<\infty\}$, which is interpreted as the
monopole worldline. We shall return to this example frequently during
the rest of the paper. 

\subsection{Magnetic spherical $p$-branes immersed in fluxbranes}

The obvious way to construct higher dimensional $p$-brane solutions is
to take the product of (\ref{monopole}) with $p$ flat spatial
directions; these will be magnetic $p$-brane solutions of a $(p+5)$-dimensional
Kaluza-Klein theory. However, an alternative construction
was presented in \cite{horowitz}, resulting in a spherical $p$-brane worldvolume. This has recently been generalised to tubular branes
\cite{refs}. We describe the case of a magnetically charged
spherical $p$-brane in a $(p+5)$-dimensional Kaluza-Klein
theory. Asymptotically these solutions all approach the fluxbrane
solutions described in references \cite{horowitz} and \cite{refs}.

The solution is again a metric product $X=\mathbb{R}\times{M}$ with
$\mathbb{R}$ a trivial time direction, but now the manifold $M$ is
the $(p+4)$-dimensional Euclidean Schwarzschild solution

\be
ds^2=-dt^2+\left(1-\left(\frac{r_H}{r}\right)^{p+1}\right)d\tau^2+\left(1-\left(\frac{r_H}{r}\right)^{p+1}\right)^{-1}dr^2+r^2d\Omega_{p+2}\label{string}\ee

where $d\Omega_n$ is the metric on the unit $n$-sphere. We write the
metric on $S^{p+2}$ as 

\be
d\Omega_{p+2}=d\theta^2+\sin^2\theta{d}\phi^2+\cos^2\theta{d}\Omega_p\ee

and take the circle action generated by the Killing vector field 

\be
\frac{\partial}{\partial\tau}+\frac{1}{R}\frac{\partial}{\partial\phi}\ee

where $R=\frac{2r_H}{p+1}$, and then Kaluza-Klein reduce on the circle. The stationary points of
the circle action are given by $\{r=r_H, \theta=0,
-\infty<t<\infty\}$, which is interpreted as the $p$-brane
worldvolume. The spatial section is the sphere $S^p$, and by analogy with the monopole solution one can see that
the $p$-brane in the base is magnetically charged under the Kaluza-Klein gauge
field. The details may be found in the original reference
\cite{horowitz}. The magnetic $p$-brane solution asymptotically approaches a fluxbrane solution at large radius.

\subsection{The Type IIA D6-brane}

Our final example is the D6-brane of Type IIA String Theory. This is
magnetically charged\footnote{D-brane charge should
properly be understood in terms of K-Theory \cite{ktheory}. Provided
one is not interested in subtleties such as the precise integrality
conditions satisfied by the charges, the use of cohomology is perfectly
adequate. In particular, it is sufficient for our purposes.} under
the Kaluza-Klein two-form that derives from the $D=11$ metric. In fact,
we have already covered this example in our comments in the last
section. One may construct, at the level of supergravity, a BPS
D6-brane. Specifically, the $M$-Theory solution is

\be
ds^2=-dt^2+dx_1^2+\ldots+dx_6^2+ds^2_{\mathrm{Taub-NUT}}\label{d6}\ee

Since the Taub-NUT solution is hyperK$\ddot{\mathrm{a}}$hler, it
admits two independent parallel spinors\footnote{covariantly constant sections of
the spin bundle.}. The D6-brane solution
(\ref{d6}) therefore breaks half of the supersymmetries of the
$M$-Theory vacuum.

\sect{Kaluza-Klein circle reduction}

In this section, we describe Kaluza-Klein reduction on a circle, where the
circle direction is given by a smooth $U(1)$ action. This action must
be \emph{semi-free} if we are to avoid orbifold singularities in the
reduced space. This is largely misunderstood in the physics literature,
so we give a very careful treatment. We illustrate the general
discussion throughout with concrete examples.

We briefly remind the reader of some definitions
regarding group actions on manifolds \cite{bredon}. A smooth action
$\Phi:G\times{M}\rightarrow{M}$ of the group $G$ on $M$ is said to be
free if given any $p\in{M}$ such that $\Phi(g,p)=p$, then $g$
is the identity element $e\in{G}$. A point $p\in{M}$ is said to be fixed if
there is some non-trivial $g\in{G}$, $g\neq{e}$, such that $\Phi(g,p)=p$. Thus we may say that
the action of $\Phi$ is free if it has no fixed points. 

A point $p\in{M}$ is said to be stationary if $\Phi(g,p)=p$,
$\forall{g}\in{G}$. We specifically make this distinction between
fixed points and stationary points as it will be important below;
often in the literature one finds that no such distinction is
made. The action of $\Phi$ is said to be semi-free if all
fixed points are in fact stationary points. It follows that the isotropy
groups $G_p\subset{G}$ are either all of $G$, or the trivial group,
$\forall{p}\in{M}$. Finally, the action is said to be effective if
each $g\neq{e}$ in $G$ moves at least one point in $M$; that is, if
$\Phi(g,p)=p$, $\forall{p}\in{M}$, then $g=e$. We tacitly assume that
our group actions are effective in this paper.

\subsection{Stationary points of circle actions}

The total spacetime manifold is $X=\mathbb{R}\times{M}$, of dimension
$D$, and is
typically a trivial metric product, although the Riemannian metric $g$
on $M$ could in principle depend on $t$. In the rest of this section, we
deal only with the space $M$; that is, the circle action on the factor
of $\mathbb{R}$ is always trivial, and thus we may neglect it.

Let $(M,g)$ be an oriented Riemannian manifold of dimension
$(d+1)$, admitting a smooth orientation-preserving isometric
circle action $\Phi_{\tau}:M\rightarrow{M}$, where $\tau$
parameterises the $U(1)$ group\footnote{In fact,
$\Phi:U(1)\times{M}\rightarrow{M}$ is smooth if and only if $\Phi_{\tau}$ is
smooth for each $\tau$ \cite{zippin}.}. Let $M^{U(1)}$ denote the set
of stationary points. Then each connected component of $M^{U(1)}$ constitutes a
closed oriented totally geodesic submanifold of $M$ of even
codimension. Let
$F$ be such a component, of codimension $2r$, and consider the induced
action of $U(1)$ on the tangent space $T_pM$, where $p\in{F}$. $T_pM$ is a real
$U(1)$-module, and hence we may decompose the $U(1)$ action into its
irreducible real representations, which, since $U(1)$ is cyclic, are
either of the form $\pm1$ or $R(\theta)=\left(\begin{array}{cc}\cos{\theta}
& -\sin{\theta}\\ \sin{\theta} &
\cos{\theta}\end{array}\right)$. Since $F$ is stationary, the
action of $\Phi_{\tau}$ on $T_pF$ is trivial, and hence we see that
the action on the normal space $N_pF$ of $F$ in $M$ may be decomposed
into the product of $r$ commuting $2\times2$ rotations in $r$
orthogonal $2$-planes. If $k$ denotes the Killing vector field
associated with $\Phi_{\tau}$, with some normalisation, then $\nabla{k}$ is a 2-form as a
consequence of Killing's equation. With respect to an orthonormal
frame at $p$, $\nabla_bk_a$ is therefore skew-symmetric, and is an
element of the Lie algebra of $SO(d+1)$. Hence with respect to the orthonormal
frame, the $U(1)$ action may be written as the direct sum
$1_{d+1-2r}\oplus\bigoplus_{j=1}^r{R}(\kappa_j\tau)$ where $\kappa_j$
are the skew eigenvalues of $\nabla_bk_a$, and the symbol $1_{d+1-2r}$
denotes the trivial action on $T_pF$. $N_pF$ then has a canonical complex structure
in which the $U(1)$ action at the tangent space level acts as
$e^{i\kappa_j\tau}\in\mathbb{C}$ in the j$^{\mathrm{th}}$ $2$-plane, which we
identify with $\mathbb{C}$. Since the action is periodic,
it follows that the skew eigenvalues $\{\kappa_j\mid{j}=1,\ldots,r\}$ must be rationally
related, the integers $\{n_j\in\mathbb{Z}\mid{j}=1,\ldots,r\}$ relating the
eigenvalues determining the number of rotations in each orthogonal
$2-$plane in $N_pF$ induced by a single orbit of the
$U(1)$ group. Since we
require the action to be effective, the
$\{n_j\mid{j}=1,\ldots,r\}$ necessarily have no common factor. Defining canonical complex
coordinates $z_1,\ldots,z_r$ on the normal space $N_pF$, the action of
$\Phi_{\tau{*}}$ is $z_j\rightarrow{e}^{in_j\tau}z_j$
for each $j$, where we have taken $\tau$ to have the canonical period
$2\pi$, and so $\kappa_j=n_j$. Of course, this discussion is
independent of the choice of point $p\in{F}$. Hence, for a generic
connected stationary point set $F$, the circle action canonically
decomposes the normal bundle $NF=\cup_{p\in{F}}N_pF$ of $F$ in $M$ into the sum of
$r$ complex line bundles, the induced action on $NF$ being
characterised by $r$ integers with highest common factor $1$.

To illustrate this discussion, let us consider the isolated stationary
point $\{r=a\}$ of the Taub-NUT metric (\ref{taubnut}). Locally, one may
choose coordinates in a neighbourhood of this point such that the
metric looks like the flat metric on $\mathbb{R}^4$, and the circle
action generated by $k=\partial/\partial\psi$ becomes the action
$z_j\rightarrow{e}^{in_j\tau}z_j$ where $n_1=n_2=(\pm)1$ and
$\{z_j\mid{j}=1,2\}$ are complex coordinates on
$\mathbb{C}^2=\mathbb{R}^4$. Thus we see that the circle action on
Taub-NUT, generated by the Killing vector field $k$, is semi-free. The
surfaces of constant $r>a$ are topologically three-spheres, and the restriction
$k\mid_{S^3}$ generates the Hopf action on $S^3$, with projection $\mathcal{H}:S^3\rightarrow{S}^2$. Alternatively,
one may take the circle action generated by $-k$. In this case
$n_1=-n_2=(\pm)1$, the resulting action on the three-spheres is the
antiHopf action, and one now has an antimonopole, rather than a
monopole. 

For an example of a non-semi-free action, simply take $\mathbb{C}^2$
with the action $z_j\rightarrow{e}^{in_j\tau}z_j$ with at least one of
the $n_j\neq\pm1$.

\subsection{Circle reduction}

We wish to perform a Kaluza-Klein circle reduction on the orbits of
$k$. In order to do this, one must form the quotient space $M/U(1)$. Since the
orbits completely degenerate on the stationary points, it is clearly
desirable to remove them before taking the quotient. We will later interpret these geodesic submanifolds as topological defects on the base space
$M/U(1)$. Now, if $M^{\prime}$ is a manifold equipped with a smooth effective action of a compact Lie group $G$ with finite isotropy
groups $G_p$ at each $p\in{M}^{\prime}$, then the quotient space $M^{\prime}/G$ has the
canonical structure of an orbifold \cite{buc}. An orbifold is a generalisation of the orbit space of a smooth effective finite group
action on a manifold\footnote{Note that this \emph{is} the definition
of an orbifold used by physicists.}. More specifically, an orbifold is a
topological space that can be covered by open sets $U_i$ homeomorphic
to $\tilde{U}_i/\Gamma_i$ where the $\Gamma_i$ are finite groups acting
smoothly and effectively on $\tilde{U}_i$, open in $\mathbb{R}^n$. In
the case at hand, since the only proper subgroups of $U(1)$ are finite
(of the form $\mathbb{Z}_m$), we see that the quotient space $M^{\prime}/U(1)$
is an orbifold, where $M^{\prime}=M-M^{U(1)}$. This fact seems to have gone
unnoticed in the physics literature.

Since we would like the base space to be a manifold, it follows that
we should only consider semi-free U(1) actions; that is, U(1) actions
whose only fixed points are stationary points. Then the isotropy
groups of $M^{\prime}$ are all trivial, and the base does indeed inherit a
genuine manifold structure. We then have a $U(1)$ principal
bundle\footnote{We denote by $M^{\prime}$ both the bundle and the
total space.}

$$\pi:M^{\prime}\rightarrow{B^{\prime}}$$

Such bundles are classified by their first Chern class
$c_1(M^{\prime})\in{H}^2(B^{\prime};\mathbb{Z})$; that is, $U(1)$ bundles over $B^{\prime}$ are,
up to isomorphism, in 1-1 correspondence with elements of the second
singular cohomology group of $B^{\prime}$ with coefficients in
$\mathbb{Z}$. One way to see this\footnote{See \cite{witten1} for a
particularly nice account of these ideas.} is to note that the classification
of $G$-bundles over $B^{\prime}$ depends on the homotopy groups $\pi_n(G)$ of
$G$, and these are all trivial for $n\geq2$ in the case that $G=U(1)$,
and $\pi_1(U(1))\cong\mathbb{Z}$. Since $B^{\prime}$ is orientable, by Poincar$\acute{\mathrm{e}}$ duality,
$H^2(B^{\prime};\mathbb{Z})\cong{H}_{d-2}(B^{\prime},\partial{B^{\prime}};\mathbb{Z})$ where
the latter denotes the $(d-2)$ relative homology group of the pair $(B^{\prime},\partial{B^{\prime}})$ and
$\mathrm{dim}(B^{\prime})=d$. Let $S$ be a codimension two submanifold of $B^{\prime}$
whose image $[S]\in{H}_{d-2}(B^{\prime},\partial{B^{\prime}};\mathbb{Z})$ is dual to
$c_1(M^{\prime})$. Then $S$ is a Dirac string, whose lift to
$M^{\prime}$ is referred to as a Misner string in \cite{hunter}. These codimension
two submanifolds are rather heuristically described in the physics
literature as submanifolds on which the foliation by surfaces of
constant $\tau$ breaks down, due to non-trivial twisting
of the $U(1)$ bundle. The term string is perhaps somewhat of a
misnomer; only when $d=3$ does one actually obtain curves of real
dimension one. We see that any codimension two submanifold $S$
defines a $U(1)$ bundle up to isomorphism, and, conversely, any $U(1)$
bundle defines $S$ up to homology. To see that $S$ is indeed an
obstruction to triviality, suppose that $S$ represents the first Chern
class of the bundle $\pi:M^{\prime}\rightarrow{B^{\prime}}$. Then if $U\subset{B^{\prime}}$ is
open, the first Chern class of the restriction $M^{\prime}\mid_{U}$ is represented by
the submanifold $S\cap{U}$ of $U$. Applying this fact
to $U=B^{\prime}-S$ implies that the first Chern class of $M^{\prime}\mid_{B^{\prime}-S}$ is represented by
$0\in{H}^2(B^{\prime}-S;\mathbb{Z})$, and therefore the restriction
$\pi:M^{\prime}\mid_{B^{\prime}-S}\rightarrow{B^{\prime}}-S$ is trivial. The singular
(co)homology theory is perhaps less familiar to physicists than the de
Rham theory, but the use of the singular theory was crucial in our
derivation above. We have
$H^2(B^{\prime};\mathbb{R})\cong{H}^2(B^{\prime};\mathbb{Z})\bigotimes_{\mathbb{Z}}\mathbb{R}\cong{H}^2_{\mathrm{dR}}(B^{\prime})$
and so the de Rham cohomology only measures the free part of
$H^2(B^{\prime};\mathbb{Z})$. The de Rham theory is therefore too crude to classify $U(1)$
bundles in general, although in many cases of interest the
torsion (the finite part of $H^2(B^{\prime};\mathbb{Z})$) vanishes and
the two approaches are equivalent. 

Let us briefly turn back to the Taub-NUT instanton (\ref{taubnut})
again to illustrate these abstract points. Taub-NUT is an
example of a space containing a Misner string. Removing the nut $\{r=a\}$ yields a
manifold of topology $\mathbb{R}^4-\{\mathrm{pt}\}$. Dividing out by
the free circle action generated by $\partial/\partial\psi$ yields a
manifold diffeomorphic to $\mathbb{R}^3-\{\mathrm{pt}\}$. The
two-sphere $S^2$ is therefore a deformation retraction of this base, and
the first Chern class of the $U(1)$ bundle is easily seen to be
$1\in\mathbb{Z}\cong{H}^2(S^2;\mathbb{Z})\cong{H}^2(\mathbb{R}^3-\{\mathrm{pt}\};\mathbb{Z})$.
The Poincar$\acute{\mathrm{e}}$ dual to the Kaluza-Klein two-form
$G_2=\frac{1}{4\pi}\sin\theta{d}\theta\wedge{d}\phi\in\Omega^2(\mathbb{R}^3-\{\mathrm{pt}\})$
may be taken to be the ray $\{\theta=0\}$. This lifts to the two-manifold
$\{\theta=0\}$ in the total space, which is therefore by
definition a Misner string of Taub-NUT space. As we proved earlier,
deleting this string trivialises the bundle. This may be seen
explicitly here. Deleting the ray $\{\theta=0\}$ from the base $B^{\prime}$
is equivalent to deleting a point from the two-sphere $S^2$ that is a
deformation retraction of $B^{\prime}$. This leaves us with
$\mathbb{R}^2$. But $H^2(\mathbb{R}^2;\mathbb{Z})\cong0$ and so by the
classification theorem, the bundle must be trivial.

\begin{figure}[t]
\vspace{1pc}
\includegraphics[-30,0][100,130]{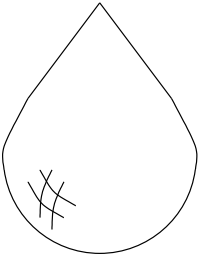}
\caption{Thurston's teardrop, $\mathbb{C}P^{[1,p]}$. A neighbourhood of the north pole is diffeomorphic to
$\mathbb{R}^2/\mathbb{Z}_p$ whereas the south pole is regular.}
\end{figure}

After this slight digression, let us now look at the removal of the stationary points in a little more
detail. Let $F$ be as above, and let $NF^{\epsilon}$ denote the open disc bundle of radius
$\epsilon>0$. This is simply defined as the space of all vectors in
$NF$ of length at most $\epsilon$. If $F$ is compact\footnote{If $F$ is non-compact, but of
course still closed as a subspace of $M$, one
must in general take $\epsilon:F\rightarrow\mathbb{R}^+$ to be a
positive function on $F$; the tubular neighbourhood theorem now goes
through. Such details will not be important, so we ignore this technicality.}, by the tubular neighbourhood
theorem one may find an $\epsilon>0$ such that the exponential map
maps $NF^{\epsilon}$ equivariantly\footnote{Recall that a map between
two $G$-spaces (spaces with a given action of the group $G$) is said
to be equivariant if it commutes with the group actions. In the case
at hand, $\exp$ is equivariant since it is defined canonically in
terms of the metric, which is $G$-invariant.} into a tubular neighbourhood of $F$
in $M$. The frontier of this neighbourhood is thus a sphere-bundle over
$F$. How does this boundary reduce under the $U(1)$ action? Let
$p\in{F}$ and consider the normal space $N_pF$. Define the sphere of
radius $\epsilon$ in $\mathbb{C}^r$ as
$S^{2r-1}_{\epsilon}=\{\{z_1,\ldots,z_r\}\in\mathbb{C}^r\mid\sum_{j=1}^r|z_j|^2=\epsilon^2\}$.
Under the action of $\Phi_{\tau{*}}$ we have $z_j\rightarrow{e}^{in_j\tau}z_j$ on
$\mathbb{C}^r$. Projecting out by this action yields a \emph{weighted projective space}, denoted
$\mathbb{C}P^{[n_1,\ldots,n_r]}$, and is a complex orbifold, of
complex dimension $r-1$, for general
$\{n_j\}$. These spaces are not uncommon in the physics
literature\footnote{Note also that $\mathbb{C}P^{[1,p]}$ is Thurston's
teardrop.}. Indeed, in \cite{candelas}, a large number of Calabi-Yau
3-folds were constructed by resolving various hypersurfaces in
$\mathbb{C}P^{[n_1,\ldots,n_5]}$. One may characterise the orbifold
points as follows. Let
$[z_1,\ldots,z_r]\in\mathbb{C}P^{[n_1,\ldots,n_r]}$ and let
$m=\mathrm{hcf}\{n_j\mid{z}_j\neq0\}$. The points with $m>1$ correspond
to orbifold points, with group $\Gamma=\mathbb{Z}_m$. The set of
regular points $X_{\mathrm{reg}}$ with
$m=1$ is dense in $\mathbb{C}P^{[n_1,\ldots,n_r]}$ and is a genuine
manifold. One should note that weighted projective spaces are not in
general global orbifolds; that is, they cannot be realised as
$Y/\Gamma$ for some manifold $Y$ and finite group $\Gamma$. The reason
we mention these facts is that they have been completely overlooked in
the physics literature. From the above discussion, we see that only in the case that
$n_j\in\{\pm1\}$ for each $j$ does one obtain a manifold, namely the familiar complex projective space $\mathbb{C}P^{r-1}$. The projection
$\mathcal{H}:S^{2r-1}\rightarrow\mathbb{C}P^{r-1}$ is then the Hopf
map (or antiHopf map, depending on orientation). Thus a necessary
condition that $B^{\prime}$ be a manifold is that all of the $n_j$, associated
with each connected component of $M^{U(1)}$, be equal to $\pm1$. We
therefore assume that the action of $\Phi$ is semi-free in the sequel. 

\sect{Brane sources and the Thom class}

In this section, we describe more precisely how the stationary point sets in
$M$ may be viewed as branes (topological defects) on the base
$B$. We reiterate that the total and reduced spacetimes are $\mathbb{R}\times{M}$
and $\mathbb{R}\times{B}$ respectively, but that we deal only with the
spatial part of the brane worldvolume in the following. Codimension
four stationary point sets in $M$ are of particular interest,
since the corresponding branes may be magnetically charged with respect to the
Kaluza-Klein two-form, $G_2$. We derive the corresponding
equation (\ref{brane}) in the cohomology group $H^3(B)$ using an explicit construction of the Thom class
$u\in{H}^3(E,E_0;\mathbb{Z})$ of the normal bundle $E$ of the brane
worldvolume $W$ in the base $B$ \footnote{note that $W$ is not currently part
of the base $B$; we shall correct this momentarily.}, where $E_0$
denotes the complement of the zero section of $E$. The brane $W$
provides a source for the Kaluza-Klein field strength
$G_2$. Specifically, the equation (\ref{brane}) implies that there is
a Wess-Zumino source term (\ref{coupling}) present in the the Kaluza-Klein reduced action\footnote{which should
be in the 'string frame', as explained later.}.This contribution to the action is familiar for example in String
Theory where the perturbative critical dimension is $D-1=10$. In this case $C_7$ is a RR-form potential under which the D6-brane is
charged \cite{polchinski}.

\subsection{Branes as stationary point sets}

Let us recapitulate our general setup. $(M,g)$ is an oriented
Riemannian manifold, admitting a smooth semi-free
orientation-preserving isometric circle action. Let $F$ denote a
codimension $2r$ connected stationary point set, which is necessarily
closed as a subspace of $M$. In order to form the
quotient space $B$, we first remove an open invariant tubular
neighbourhood $NF^{\epsilon}$ around $F$, yielding the space $M^{\prime}$. The limit
in which the radius of this neighbourhood goes to zero corresponds to
just removing the stationary points $F$. If $F$ is compact, the radius
may be taken to be $\epsilon>0$ constant. Otherwise, one may have to take the
radius to be a function on $F$;
$\epsilon:F\rightarrow\mathbb{R}^+$. The frontier of the tubular neighbourhood is a $(2r-1)$-sphere bundle over $F$, which is a deformation retraction of the
complement of the zero section $NF_0$ of the normal bundle $NF$ of $F$
in $M$. The circle action simply corresponds to moving along the
fibres of the Hopf (or anti-Hopf) fibration of the $(2r-1)$-sphere,
where we have identified $NF^{\epsilon}$ equivariantly with a tubular
neighbourhood of $F$ in $M$ via the exponential map. 

Now, the image
of this frontier in the base $B^{\prime}=M^{\prime}/U(1)$ is a
$\mathbb{C}P^{r-1}$ bundle over $F$. Now we come to an important
point. This boundary may be interpreted as a brane in $B^{\prime}$. The case
$r=2$ is special since the boundary in $B^{\prime}$ is an
$S^2=\mathbb{C}P^1$ bundle over $F$, which we may 'fill in' by glueing it to the boundary of
an appropriate oriented closed disc bundle over $F$. More precisely,
the transition functions for this closed disc bundle are
given by the transition functions of the two-sphere bundle boundary, so that the
latter may be regarded as the boundary of the disc bundle over $F$. The zero section is
interpreted as the brane worldvolume $W$, and is diffeomorphic to $F$
(although we use different names, to distinguish logically between the
submanifold $F$ of $M$ and the image $W$ in the base $B$). We call the
resulting space $B$, which now has no boundary associated with $F$. We
also have a projection $\pi:M\rightarrow{B}$ where the image of $F$ is $W$,
and the restriction to $M^{\prime}$ is a $U(1)$ bundle. This
construction \emph{only} works for codimension four. The simple reason
is that a $\mathbb{C}P^{r-1}$ bundle over some space $W$ is a deformation
retraction of the complement of the zero section of a vector bundle over
$W$ only if $r=2$ (the case $r=1$ is rather trivial as far as we are
concerned)\footnote{In fact, $\mathbb{C}P^{2n}$ is not the oriented boundary of
any oriented $(4n+1)$-manifold with boundary, for
$n\geq1$. We shall have more to say on
cobordism in Section 5.}. Euclidean space foliates into spheres, not complex
projective spaces. The case at hand, however, is degenerate since $S^2=\mathbb{C}P^1$. This is interesting, since branes in $B$ can be magnetically charged with
respect to the Kaluza-Klein two-form $G_2$ precisely in this dimension. We now show
that this is indeed the case. The construction above leads
to the interpretation of $W$ as a magnetically charged brane in the
base $B$, satisfying equation (\ref{brane}). However, before proving
this, we first remind the reader of some facts about branes.

\subsection{Brane sources}

Let us recall the general theory of a $(p-1)$-form potential $C_{p-1}$,
with field strength $G_p=dC_{p-1}$, on a $(D-1)$-dimensional spacetime
$Y=\mathbb{R}\times{B}$, with $\mathbb{R}$ a trivial time direction. We
assume that $B$ contains codimension (p+1) branes $W$ that are magnetically charged with
respect to $C_{p-1}$. Specifically, one has a Wess-Zumino source term

\be
\pm(-1)^{D-p}\int_{\mathbb{R}\times{W}}C_{D-p-2}\ee

with a $+$ for branes and a $-$ for antibranes, together with the
usual bulk action which should be in the 'string' frame

\be
-\frac{1}{2}\int_Y*G_p\wedge{G}_p\ee

where $C_{D-p-2}$ is the
potential for the Hodge dual field strength
$dC_{D-p-2}=G_{D-p-1}=*G_p$, the Hodge dual being that defined by the
\emph{Lorentz} metric on $\mathbb{R}\times{B}$ (for example, just take a
trivial metric product. The details are not too important). Varying
this action leads to the equation in cohomology

\be
[dG_p]=\pm[\delta(W)]\label{genbrane}\ee

where $\delta(W)$ is a closed $(p+1)$-form that is the Poincar$\acute{\mathrm{e}}$ dual
of $W$ in $B$, and whose support is limited to $W$, and $[\ldots]$ denotes the
image in $H^*(B;\mathbb{R})$. 

 The equation
(\ref{genbrane}) follows from demanding stationarity of the action under
the variation $C_{D-p-2}\rightarrow{C}_{D-p-2}+\delta{C}_{D-p-2}$,
such that the cohomology class of $G_{D-p-1}$ is preserved. Note that
the variation of the Wess-Zumino term gives

\be
\pm(-1)^{D-p}\int_{\mathbb{R}\times{W}}\delta{C}_{D-p-2}\ee

Since $\delta{C}_{D-p-2}$ is closed, we may rewrite the integral as

\be
\int_{\mathbb{R}\times{W}}\delta{C}_{D-p-2}=\int_Y\delta{C}_{D-p-2}\wedge\eta_{p+1}\ee

where $\eta_{p+1}$ is a closed $(p+1)$-form whose cohomology class is
Poincar$\acute{\mathrm{e}}$ dual to $W$ in $B$ (see, for example,
\cite{bott}).

Varying the whole action therefore gives

\be
-\frac{1}{2}\delta(*G_p\wedge{G}_p)+\delta(\pm(-1)^{D-p}C_{D-p-2}\wedge\eta_{p+1})=0\ee

and so

\be
-\frac{1}{2}\delta(d{C}_{D-p-2})\wedge{G}_p-\frac{1}{2}(-1)^{p(D-p-1)+1}*G_p\wedge*\delta(dC_{D-p-2})+\pm(-1)^{D-p}\delta{C}_{D-p-2}\wedge\eta_{p+1}=0\ee

We have used $*^2=(-1)^{p(D-p-1)+1}$ on $p$-forms. Since $d$ commutes with $\delta$, we have

\be
\delta{C}_{D-p-2}\wedge{d}G_p=\pm\delta{C}_{D-p-2}\wedge\eta_{p+1}+d\lambda\ee

where $\lambda$ is a global\footnote{This word is crucial. Although
$C_{D-p-2}$ is not a global form in general, its variation
$\delta{C}_{D-p-2}$ \emph{is} global.} $(D-2)$-form on $Y$. Hence, taking the
image in cohomology, and
noting\footnote{This is refered to as the Localisation Principal.} that
the support of the Poincar$\acute{\mathrm{e}}$ dual $[\eta]$ of a closed submanifold $W$ may be shrunk
into any tubular neighbourhood of $W$, we finally obtain equation
(\ref{genbrane}), in the formal limit that the support is shrunk onto
$W$ yielding a delta-function $\delta(W)$. For a general set of branes
$\{W\}$ and antibranes $\{\bar{W}\}$, one may extend the above argument
linearly to yield

\be
[dG_p]=\sum[\delta(W)]-\sum[\delta(\bar{W})]\label{genbrane2}\ee

as an equation in $H^{p+1}(B)$. If $W$ is compact, the
Poincar$\acute{\mathrm{e}}$ dual of $W$ has compact support. Provided
that the variation $\delta{C}_{D-p-2}$ is assumed to have compact
support, we see that equation
(\ref{genbrane}) also holds in the compactly supported cohomology group
$H^{p+1}_c(B)$, defined in the next section. This will be important
for our definition of brane charge. Note that we made the somewhat peculiar choice of coupling
$(-1)^{D-p}\int_{\mathbb{R}\times{W}}C_{D-p-2}$ precisely in order that equation (\ref{genbrane})
has no dependence on dimension. This is natural, as we shall
now show from the Kaluza-Klein perspective.

\subsection{The Thom class}

We now derive equation (\ref{genbrane}) from our Kaluza-Klein
perspective, where $p=2$. In this case, $G_2$ is the Kaluza-Klein
field strength. Appropriately normalised, its image in
cohomology is therefore the first Chern
class $c_1(M^{\prime})\in{H}^2(B^{\prime};\mathbb{Z})$. We assume that the (not necessarily connected)
stationary point set $M^{U(1)}$ yields a configuration of branes $\{W\}$
and antibranes $\{\bar{W}\}$, arising from each connected component $F$ of
$M^{U(1)}$ via the construction in Section (4.1). More specifically,
recall that the normal bundle of $F$ in $M$ is a rank four oriented
vector bundle, its orientation being the canonical one\footnote{that
is, such that $NF\oplus{T}F=TM\mid_F$ has the direct sum
orientation.}. Then, as in Section 3, for a semi-free circle action,
the induced action on $NF$ is characterised by two integers
$n_1,n_2\in\{\pm1\}$. If $n_1=n_2$ then we interpret $F$ as a brane,
otherwise $F$ is an antibrane. We could of course have chosen this the
other way around, or, alternatively, we could change the orientation of
$NF$. The point is that the choice is unphysical. Interchanging branes
with antibranes is a symmetry of the theory; only the relative sign is
important. We denote the disjoint
sum of branes and antibranes in $B$ as
$\mathcal{W}=\mathcal{W}^++\mathcal{W}^-$, separating $\mathcal{W}$ into its brane and antibrane constituents.

Let $p:E=N\mathcal{W}\rightarrow{\mathcal{W}}$ denote the normal bundle of $\mathcal{W}$ in the
\emph{base} $B$. This is a rank three orientable vector bundle over
$\mathcal{W}$. The Hopf map
$\mathcal{H}:S^3\rightarrow{S}^2$ has first Chern class corresponding to
$1\in\mathbb{Z}\cong{H}^2(S^2;\mathbb{Z})$, whereas the antiHopf map
corresponds to $-1$. Thus one has a brane $W\subset\mathcal{W}^+$ or
an antibrane $\bar{W}\subset\mathcal{W}^-$ depending on the sign of $G_2$. Note that flipping the sign of $G_2$
flips the sign of the coupling in (\ref{coupling}), changing a brane
into an antibrane. 

In order to derive (\ref{genbrane}), we must introduce the notion of
the Thom class of an oriented vector bundle. However, before doing
this, we first remind the reader of some definitions. We have assumed
so far that the reader is familiar with cohomology groups. For de Rham
cohomology, these are roughly speaking the space of closed forms modulo the
space of exact forms. However, on a non-compact manifold, one may also
define cohomology with compact support. In this case, the cohomology
groups are the space of closed forms with compact support modulo the
space of exact forms $d\omega$, where $\omega$ has compact
support. Thus, given a closed compactly supported $p$-form $\nu$ on
some manifold $B$, we may consider $\nu$ as an element of both $H^p(B)$ and
$H^p_c(B)$. If $\nu=d\omega$ for some global form $\omega$ then $\nu$
is trivial as an element of $H^p(B)$, but it may not be trivial as an
element of $H^p_c(B)$ since $\omega$ may not have compact
support. This will be important in the following. Finally, for forms
defined on the total space of some vector bundle $E$, we have the
notion of cohomology with compact support in the vertical
direction; in other words, the forms above need not have compact support on
$E$, but instead the restriction to each fibre is required to have
compact support. The cohomology is denoted
$H^*_{\mathrm{cv}}(E)$. Similar definitions exist for the singular
theory and may be found in Appendix A of \cite{milnor}.

Let $\pi:E\rightarrow{B}$ be a rank $n$
oriented vector bundle over $B$. Then the cohomology group
$H^n(E,E_0;\mathbb{Z})$ contains precisely one cohomology class $u$,
the Thom class, whose restriction
$u\mid_{(V,V_0)}\in{H}^n(V,V_0;\mathbb{Z})$ is the preferred generator
given by the orientation of the bundle, for each fibre
$V\cong\mathbb{R}^n$, where a subscript $0$ denotes the complement of
zero (or zero section in the case of bundles). This class enters into the Thom Isomorphism
Theorem. This states that

\be
\mathcal{T}:H^*(B;\mathbb{Z})\rightarrow{H}^{*+n}(E,E_0;\mathbb{Z})\label{thom}\ee

is an isomorphism, given explicitly by

\be
\mathcal{T}(\omega)=(\pi^*\omega)\cup{u}\label{thom2}\ee

where $\cup$ is the cup product\footnote{which should of course be
replaced by the wedge product $\wedge$ in the de Rham category.}. For
a proof of these statements, the reader is again referred to \cite{milnor}.

Alternatively, and equivalently, one may replace the relative cohomology $H^*(E,E_0)$
with cohomology with compact support in the vertical direction,
$H^*_{\mathrm{cv}}(E)$. Then the Thom class, as above, is uniquely
characterised as the cohomology class in $H^n_{\mathrm{cv}}(E)$ which restricts
to the preferred generator of $H^n_{\mathrm{c}}(V)$ on each fibre $V$, where
$H^*_{\mathrm{c}}(V)$ denotes the cohomology ring of $V$ with compact
support. It is this notion of the Thom class that we shall need.

The first point to note is that the Poincar$\acute{\mathrm{e}}$ dual of a closed
oriented submanifold $\mathcal{W}$ of $B$ and the Thom class of the normal
bundle of $\mathcal{W}$ in $B$ can be represented by the same forms. Thus if $u\in{H}^3_{\mathrm{cv}}(T)$
is the Thom class of a tubular neighbourhood\footnote{By definition,
the exponential map is a homeomorphism onto its image here, and so the
tubular neighbourhoods around each connected component of
$\mathcal{W}$ are non-intersecting.} $T$ of $\mathcal{W}$ in $B$, and $\eta$
is the Poincar$\acute{\mathrm{e}}$ dual of $\mathcal{W}$, we have 

\be
\eta=j_*u\ee

as a relation in $H^3(B)$, where $j_*$ denotes extension of $u$ by
zero. The proof is straightforward; one merely shows that $j_*u$
satisfies the defining equation of the Poincar$\acute{\mathrm{e}}$
dual.

Alternatively, in terms of the relative theory, since $T$ and $B-\mathcal{W}$ have union $B$ and intersection
$T-\mathcal{W}$, one has an excision isomorphism\footnote{See, for example,
\cite{maunder}. Roughly speaking, the idea is that excising simplexes
in $B-T$ from both $B$ and $B-W$ does not alter the (co)homology.} 

\be
i^*:H^*(B,B-\mathcal{W})\rightarrow{H}^*(T,T-\mathcal{W})\ee

where $i$ denotes inclusion. Thus, combining $(i^*)^{-1}$ with the
restriction map $H^*(B,B-\mathcal{W})\rightarrow{H}^*(B)$ maps the Thom class $u\in{H}^3(T,T-\mathcal{W})$ to
the Poincar$\acute{\mathrm{e}}$ dual of $\mathcal{W}$, $\eta\in{H}^3(B)$.

Going back to our general discussion of the Thom class of an oriented
vector bundle $E$, we now
describe an explicit construction for a representative of the Thom class in terms of the
global angular form $\psi$ on $E_0$. This is described in
\cite{bott}. It is essentially the vector bundle analogue of passing
from a generator of $H^{n-1}(S^{n-1})$ to a generator of
$H^n_{\mathrm{c}}(\mathbb{R}^n)$. Given an Euclidean vector bundle
$E$, one may define an associated sphere bundle $S(E)$ given by the
subbundle consisting of all unit vectors in $E$. Then one has a
deformation retraction $f:E_0\rightarrow{S}(E)$ of the complement of the
zero section of $E$ onto $S(E)$. The global angular form
$\psi_S\in\Omega^{n-1}(S)$ on an oriented $(n-1)$-sphere
bundle has two defining properties 

\begin{itemize}

\item Its restriction to each fibre
generates the cohomology of the fibre

\item $d\psi_S=-\Pi^*e$

\end{itemize}

where $\Pi$ is the sphere-bundle projection, and $e$ is the Euler
class of the sphere bundle. When $S$ derives from a vector bundle $E$, one
may pull back $\psi_S$ to $E_0$ via the deformation $f$; thus
$\psi=f^*\psi_S\in\Omega^{n-1}(E_0)$, which we call the global angular
form of $E_0$. It is now a simple matter to prove that the
cohomology class

\be
u=[d(\rho(r).\psi)]_{\mathrm{cv}}\in{H}^n_{\mathrm{cv}}(E)\label{thomclass}\ee

is the Thom class of $E$, where $r$ is the radius (defined by the
metric) and all that is
required of the function $\rho(r)$ is that $\rho$ be smooth and equal to $-1$ in a neighbourhood of $0$, and equal to $0$ at infinity. Its
derivative $\rho^{\prime}(r)$ is then typically a bump function on
$\{r\in\mathbb{R}\mid{r}\geq0\}$ with total integral $1$. Although
$\psi$ is defined only on $E_0$, we have $d\rho\equiv0$ in a neighbourhood
of $0$, from which it follows that $d(\rho(r).\psi)$ is a global form
on $E$. One must show that this form is closed (which is trivial),
has compact support in the vertical direction, and has unit integral along
each fibre. The last two conditions follow from the defining
properties of $\rho(r)$ and $\psi$. Also, it is easy to see that any other function $\tilde{\rho}$ which has the same defining
properties as $\rho$ above yields the same cohomology class for
$u$. This completes our description of the Thom class in terms of the global
angular form.

\begin{figure}[t]
\vspace{1pc}
\includegraphics[-30,0][100,130]{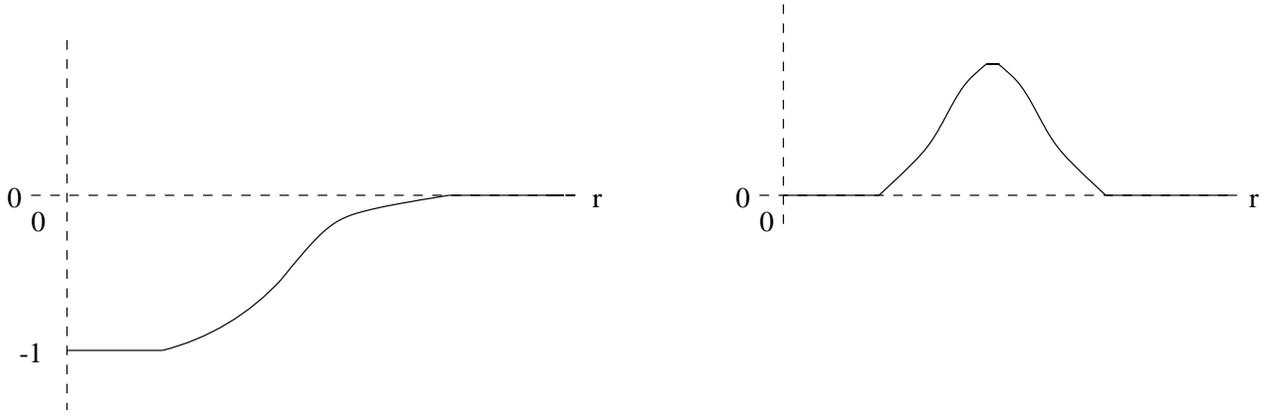}
\caption{\emph{Left}: The function $\rho(r)$. \emph{Right}: its
derivative $\rho^{\prime}(r)$ is a bump function with total integral $1$.}
\end{figure}

After this brief summary, we now turn back to our main
discussion. Consider our tubular neighbourhood $T$ of $\mathcal{W}$ in $B$. Since its rank is
odd-dimensional, the Euler class vanishes\footnote{Proof: any
odd-dimensional vector bundle admits an orientation-reversing
automorphism, and $e$ changes sign under orientation reversal.}, and
hence the global angular form $\psi$ of each connected component of $T_0$ is
closed, $d\psi=0$. Our first task is to show that $\psi=\pm{G}_2$,
depending on whether the connected component corresponds to a brane or an
antibrane, where recall that we normalise $G_2$ such that its image in cohomology
generates the first Chern class $c_1(M^{\prime})$, and we take
$M^{\prime}=M-M^{U(1)}$. Note first that $G_2$ is indeed closed on $T_0$. Since two forms are equal if and only if they are
equal locally, we may reduce the problem to a local one. Lift $T$ to
its image in $M$; this is an invariant tubular neighbourhood of $M^{U(1)}$ in
$M$ and has rank $4$. Now use the metric on $M$ to define normal
coordinates on a connected component of this
bundle. Specifically, in the case of a brane, in normal coordinates we have

\be
ds^2=\left[dr^2+\frac{r^2}{4}\left((d\alpha-\cos\beta{d}\gamma)^2+d\beta^2+\sin^2\beta{d}\gamma^2\right)\right]+dx_1^2+\ldots+dx^2_{d-3}\label{normal}\ee

The piece in square brackets is the metric restricted to the $\mathbb{R}^4$
fibres, and the remaining piece is the restriction of the metric to an
open set $U\subset{F}$. $(\alpha,\beta,\gamma)$ are Euler angles on
the unit $S^3$. The circle action simply rotates around the fibres of
the Hopf fibration of the three-spheres, which are themselves fibrered
over $F$. It follows that, in this coordinate system,

\be
G_2=\frac{1}{4\pi}d(-\cos\beta{d}\gamma)=\frac{1}{4\pi}\sin\beta{d}\beta\wedge{d}\gamma\ee

This indeed integrates to $1$ on each $S^2$ fibre, which is as expected,
since the image in cohomology of $G_2$ restricted to an $S^2$ fibre
is $c_1(\mathcal{H}:S^3\rightarrow{S}^2)\in{H}^2(S^2;\mathbb{Z})\cong\mathbb{Z}$ which is
$1\in\mathbb{Z}$ for the Hopf bundle. Hence $G_2$ generates the
cohomology of the unit two-sphere fibres. For the case of an antibrane $\bar{W}$, one has
the antiHopf, rather than Hopf, action. Hence the cohomology class of
$G_2$ in $H^2(S^2;\mathbb{Z})$ is $-1$, and $-G_2$ now generates the
cohomology of the fibres. It follows that $\psi=\pm{G}_2$ holds in a tubular
neighbourhood of a brane or antibrane, respectively.

For the sake of exposition, we focus on the case of a
single compact brane $W$. The more general case of an arbitrary set of
compact branes and antibranes follows in a straightforward manner. Note that $G_2$ is
not currently defined on the whole space $B$. In order to prove
equation (\ref{genbrane}) as a relation in $H^3_c(B)$, we first
construct an extension of $G_2$ on the whole of $B$, and then show that
(\ref{genbrane}) is independent of the choice of extension. 

In $T_0$ we have

\be
0=dG_2=d(1.G_2)=-d(\rho(r).G_2)+d((1+\rho(r)).G_2)\label{eq}\ee

where $\rho(r)$ is the function defined above,
with the additional requirement that $\rho\equiv0$ in a neighbourhood
of the boundary of $T$. It follows that $d(\rho(r).G_2)$
and $d\nu$ are equal as forms on $T_0$, where
$\nu=(1+\rho(r)).G_2$. Now, since $G_2=\psi$ for a brane, it follows
that the first term on the right hand side of (\ref{eq}) is a
representative of the Thom
class $u$ of the normal bundle of $W$ in $B$. As we remarked earlier, this is a global
form on $T$. Similarly, $\nu$ is a global form on $T$ since
the function $(1+\rho(r))$ vanishes in a neighbourhood of $r=0$, and therefore $\nu$
vanishes in a neighbourhood of the zero section. Thus,
since both forms are smooth and equal on $T_0$, we
deduce that they are equal on $T$ also (indeed, both forms vanish on
the zero section). Taking the limit in which the support
of $\rho(r)$ is shrunk to the origin $r=0$, we have $\nu=G_2$ on
$T_0$. Thus, extending the definition of $G_2$ to $\nu$ on the whole
of $T$, and taking the image in $H^3_{\mathrm{cv}}(T)$, we obtain

\be
u=[dG_2]_{\mathrm{cv}}\label{int}\ee

Extending the Thom class by zero outside $T$, we deduce that
for $W$ compact

\be
[dG_2]_c=[\delta(W)]_c\label{dG2a}\ee

holds as a relation in $H^3_c(B)$, and therefore as a relation in
$H^3(B)$. Having proven existence of an extension of $G_2$ such that
(\ref{dG2a}) holds, we now prove uniqueness. The definition of the
form $\nu$, the extension of $G_2$, is certainly not
unique. However, suppose that $\eta\in\Omega^2(T)$ is another global
extension. Then, by definition, $\eta\equiv{G}_2$ in a neighbourhood
of the boundary of $T$. Thus the form $\eta-\nu$ has compact support
in the vertical direction. Moreover, since $\eta$ is
assumed to be a global form on $T$, it follows that $\eta-\nu$ is a
global form on $T$, and thus $\eta-\nu\in\Omega^2_{\mathrm{cv}}(T)$. Hence the cohomology class of $d(\eta-\nu)$ in
$H^3_{\mathrm{cv}}(T)$ is trivial. This completes the proof. Note, in
particular, that the choice of extension of $G_2$ in (\ref{dG2a}) is independent
of the choice of function $\rho(r)$. We shall analyse equation
(\ref{dG2a}) in the context of brane charge in the next section.

For an antibrane, one merely inserts $G_2=-\psi$ into (\ref{eq}),
yielding a minus sign in (\ref{dG2a}). Hence, for a general
configuration of branes and antibranes $\mathcal{W}$, we recover
equation (\ref{genbrane2}) for the case $p=2$.

This is precisely what we wanted to show. Let us follow through the
logic. A codimension four stationary point set in a $(d+1)$-dimensional
manifold $M$ equipped with a semi-free circle action may be regarded, via the
construction in section (4.1), as a codimension three brane $W$ in the base
$d$-manifold. If $G_2$ is the Kaluza-Klein field strength, then this
brane acts as a source for $G_2$ via equation (\ref{dG2a}). Going over
to the general theory of branes, this implies that the brane must
couple to the dual potential as $(-1)^D\int_{\mathbb{R}\times{W}}C_{D-4}$; hence one must
include this together with the usual Kaluza-Klein reduced action in
order to reproduce the correct equation for $G_2$ in the presence of
the brane.

In conclusion, codimension four stationary point sets of semi-free
circle actions may be interpreted, upon Kaluza-Klein reduction, as
branes or antibranes in the base space that are magnetically charged
with respect to the Kaluza-Klein two-form. We have thus both formalised and generalised the ideas in
papers such as \cite{horowitz}.

\subsection{Brane charge}

For completeness, we describe the interpretation of equation
(\ref{genbrane2}) in terms of brane charge, as measured at
infinity. This discussion largely follows \cite{witten2}, although we
use the examples in Section 2 to illustrate the ideas, and also relate
this definition of brane charge to the usual definition in terms of
integrals of $G_2$ over two-spheres. The cohomological definition of
brane charge has considerable advantage
over the latter definition in that it unambiguously defines the total brane
charge of any configuration.

Equation (\ref{genbrane2}) states that the total brane charge, defined
to be the right hand side, is necessarily zero in the cohomology group
$H^{p+1}(B;\mathbb{Z})$, since $G_p$ is required to be globally
defined. In particular, if $B$ is compact, this is the familiar
statement that the total charge of an abelian gauge symmetry must be
zero on a compact manifold. However, in the case that $B$ has
non-empty boundary
$\partial{B}$ 'at infinity', and $W$ is compact, one may use the compact
Poincar$\acute{\mathrm{e}}$ dual
$[\delta(W)]_{\mathrm{c}}\in{H}^{p+1}_{\mathrm{c}}(B;\mathbb{Z})$, as opposed to
the closed Poincar$\acute{\mathrm{e}}$ dual
$[\delta(W)]\in{H}^{p+1}(B;\mathbb{Z})$ that we have been using in
most of the discussion so far\footnote{For a detailed account, see \cite{bott}.}. There is of
course no distinction between the two when $B$ is compact. Although
$[\delta(W)]$ is trivial in $H^{p+1}(B;\mathbb{Z})$, it is not
necessarily true that $[\delta(W)]_c$ is trivial as an element of
$H^{p+1}_{\mathrm{c}}(B;\mathbb{Z})$, the reason being that $G_p$ need
not have compact support. The interpretation given in \cite{witten2}
is therefore that brane charge should be interpreted as an element of
$H^{p+1}_{\mathrm{c}}(B;\mathbb{Z})$ whose image in
$H^{p+1}(B;\mathbb{Z})$ is trivial under the 'forgetful map'
$f:H^{p+1}_{\mathrm{c}}(B;\mathbb{Z})\rightarrow{H}^{p+1}(B;\mathbb{Z})$,
that 'forgets' that a class has compact support.

In order to interpret this definition of brane charge in terms of
fields measured at infinity $\partial{B}$, one may use the exact cohomology
sequence for the pair $(B,\partial{B})$ \cite{maunder}

\be
\ldots{H}^p(B;\mathbb{Z})\longrightarrow^{i^*}H^p(\partial{B;\mathbb{Z}})\longrightarrow^{\delta^*}H^{p+1}(B,\partial{B};\mathbb{Z})\longrightarrow^{j^*}H^{p+1}(B;\mathbb{Z})\longrightarrow\ldots\ee

where $i:\partial{B}\rightarrow{B}$ is the inclusion map, and
$j:C(B)\rightarrow{C}(B,\partial{B})$ is the quotient chain map, and,
as earlier, the relative cohomology group $H^{p+1}(B,\partial{B};\mathbb{Z})$ is
the same as the compactly supported cohomology group
$H^{p+1}_{\mathrm{c}}(B;\mathbb{Z})$ in this case. Using exactness of the
long cohomology sequence, we
thus conclude that $\mathrm{ker}(f)=H^p(\partial{B};\mathbb{Z})/i^*(H^p(B;\mathbb{Z}))$, and
so we may interpret brane charge in terms of a field strength $G_p$ on
$\partial{B}$ that does not extend over $B$ as a \emph{closed} form, and
therefore must have been created by branes. 

In the Kaluza-Klein case, note that if $G_2$ has compact support, then
$G_2$ vanishes in a neighbourhood of the boundary $\partial{B}$. By
the classification theorem, this implies that the circle bundle is
trivial in a neighbourhood of infinity. We conclude that the configuration has zero brane charge.

We now use the examples in Section 2 to illustrate these rather
abstract topological ideas.

\emph{The monopole}

In this case, the base
$B=\mathbb{R}^3$, and the monopole worldline $W$ is just a point
$x\in\mathbb{R}^3$, which we may take to be the origin. Now
$H^3(\mathbb{R}^3;\mathbb{Z})\cong0$, and we may take any three-form on $\mathbb{R}^3$ as
the closed Poincar$\acute{\mathrm{e}}$ dual of $x$. Shrinking the support of
this three-form into a neighbourhood of $x$ gives us the unit-integral
bump form

\be
f(x^1,x^2,x^3)dx^1\wedge{d}x^2\wedge{d}x^3\label{bump}\ee

where $(x^1,x^2,x^3)$ are canonical coordinates on $\mathbb{R}^3$
and $f$ is a bump function on $\mathbb{R}^3$ with support in a
neighbourhood of $x$. In the limit in which we shrink the support to
be only at the point $x$, the function $f$ becomes a Dirac delta
function, $\delta(x)$, with support at $x$.

However, the compactly supported cohomology
$H^3_c(\mathbb{R}^3;\mathbb{Z})\cong\mathbb{Z}$ is non-trivial; one may take the
unit bump form (\ref{bump}) to be one of the generators. Hence the compact
Poincar$\acute{\mathrm{e}}$ dual of the point $x$ is
non-trivial as an element of $H^3_c(\mathbb{R}^3;\mathbb{Z})$. Above, we interpreted brane charge as an element of
$H^3_c(\mathbb{R}^3;\mathbb{Z})$ that is trivial when mapped to
$H^3(\mathbb{R}^3;\mathbb{Z})$ under the forgetful map. In this case, the
Kaluza-Klein monopole charge is precisely the unit bump form (\ref{bump})
that generates $H^3_c(\mathbb{R}^3;\mathbb{Z})$, since any three-form
is trivial when regarded as an element of
$H^3(\mathbb{R}^3;\mathbb{Z})$. Having identified the monopole charge
with one choice of generator of the compactly supported cohomology,
the antimonopole charge corresponds to the other choice of generator.

\emph{The D6-brane}

The discussion for the D6-brane is straightforward. We first wrap the
directions transverse to the brane in order that the worldvolume $W$
be compact. The solution therefore takes the form (\ref{d6}) but with
$x_1,\ldots,x_6$ periodically identified. $W$ is now a six-torus,
$T^6$, located at $\{r=a\}$ in the Taub-NUT part of the spatial
section $M$. The normal bundle of $W$ in $M$ is just $\mathbb{R}^3$,
and therefore the discussion of charge reduces precisely to the case
of the monopole above. That is, the $D6$-brane
charge\footnote{modulo $K$-theoretic considerations} corresponds to a
choice of generator of
$H^3_c(\mathbb{R}^3;\mathbb{Z})\cong\mathbb{Z}$; the other choice of
generator is then associated with the charge of the
anti-D6-brane, $\bar{D6}$. 

\subsection{Relation to the usual definition of brane charge}

One would normally define the magnetic charge of a Kaluza-Klein brane $W$ to be 

\be
Q=\int_{S^2}G_2\ee

where $S^2$ is a two-sphere that surrounds the brane worldvolume
$W$. One may derive this from our discussion above by evaluating
equation (\ref{int}) on a typical fibre $V$ of a tubular neighbourhood of
$W$ in $B$. Thus, since the Thom class has unit integral on each
fibre, we deduce that

\be
Q=Qu[V]=[dG_2]_{\mathrm{cv}}[V]=\int_VdG_2=\int_{\partial{V}}G_2\ee

where $\partial{V}$ is a two-sphere that bounds the fibre
$V$, and we have reinstated the factor of $Q$ in (\ref{int}). However, the advantage of the cohomological definition of brane
charge is that it gives an unambiguous definition of the \emph{total}
brane charge of an arbitrary configuration. The problem is that one
cannot define the total brane charge of an arbitrary configuration as
an integral of the Kaluza-Klein two-form over the sphere at infinity,
since in general dimension, the sphere at infinity is not a two-sphere. Moreover, there is
no natural definition of a two-sphere at infinity in general. One can only use
this naive definition of brane charge when the base
has dimension three (for example, the monopole). In this case, the
cohomological and naive definitions agree on evaluating the former on
the fundamental homology class. Similar remarks apply to $p$-brane charge in
general.

\sect{Kaluza-Klein monopoles}

\subsection{The general setup}

In this section, we study monopole-antimonopole production in a
five-dimensional \emph{Euclidean} Kaluza-Klein theory, specialising the above
discussion to the case $d=4$. In dimension four, monopoles are
$0$-branes that are magnetically charged under the Kaluza-Klein field
strength, $G_2$. The basic static Lorentzian solution is given by
(\ref{monopole}). It turns out
that one may use various $G$-index theorems \cite{atiyah} in order to
relate the numbers and types of defects nucleated to the topology of
the defects (which is trivial here) and to the topology of $M$ (or
rather, a nucleation surface $\Sigma$ in $M$). 

The setup we consider is the following. Let $M$ be an oriented
Riemannian $5$-manifold with compact boundary $\partial{M}=\Sigma$ equipped with a smooth
isometric circle action. Since $\Sigma$ is an oriented boundary, by definition
it has trivial oriented cobordism class. By a theorem of Thom \cite{milnor},
the Hirzebruch signature $\mathrm{Sign}(\Sigma)$ of $\Sigma$ is then necessarily
zero. Here, $\mathrm{Sign}(\Sigma)$ is defined as the signature of the
non-degenerate quadratic form on $H_2(\Sigma;\mathbb{R})$ defined by
the cup-product. It is also given by
the index of a certain elliptic operator, associated with the de Rham
complex \cite{atiyah} (paper III). It also follows that the Pontrjagin
numbers and Stiefel-Whitney numbers of $\Sigma$ must all vanish
\cite{milnor}. The Euler class of $\Sigma$, $\chi(\Sigma)$ is usually defined
as the alternating sum of the Betti numbers, $b_i$, of $\Sigma$. Thus
$\chi(\Sigma)=\sum_{i=0}^{4}(-1)^ib_i$ where
$b_i=\mathrm{dim}(H_i(\Sigma;\mathbb{R}))$. However, it is also the
index of the de Rham complex. We will have more to say on this in the
next section. Since $\Sigma$ is a boundary, $\chi(\Sigma)$ is
even. To see this, we define the double of $M$ as
$2M=M\cup_{\partial{M}}(-M)$ where $-M$ denotes $M$ with its orientation reversed,
and the $\cup_{\partial{M}}$ symbol denotes that $M$ and $-M$ are to be glued
together across $\partial{M}$. $2M$ is a closed oriented manifold with
a smooth $U(1)$ action. One has the following formula for the Euler
characteristic \cite{maunder} 

\be
\chi(\partial{M})=2\chi(M)-\chi(2M)
\ee

Since $2M$ is closed and of odd dimension, $\chi(2M)=0$ trivially, and
thus we see that $\chi(\partial{M})=2\chi(\Sigma)\in2\mathbb{Z}$. This will be important in the sequel.

We want to interpret $\Sigma$ as a nucleation
surface; that is, $M$ is joined to the post-tunnelling
Lorentzian manifold $M_L$ across the totally geodesic (zero-momentum)
surface $\Sigma$. In $M_L$ $\Sigma$ is a spacelike boundary, which
serves as a Cauchy surface for $M_L$, and may be interpreted as 'the
beginning of time' \cite{hartle}. The circle isometry of $M$
should extend to a circle subgroup of the isometry group of $M_L$. We
will not consider the Lorentzian sector in detail in the rest of this paper, so we
leave out the details of precisely how the analytic continuation is to be
performed.

We now suppose that the semi-free circle action on $M$ has a
one-dimensional stationary point set, $\mathrm{dim}(M^{U(1)})=1$ (note
this need not be connected). Thus
$\nabla_bk_a$ has rank four. The stationary points then constitute oriented geodesic curves in
$M$. Since $k$ is tangent to $\Sigma$, these geodesic curves intersect
$\Sigma$ orthogonally, if at all. In this case a generic connected geodesic curve starts
on $\Sigma$, continues into $M$, and then re-intersects $\Sigma$ at some
parameter distance along the curve. These points on $\Sigma$ are the
zeros of $k\mid_{\Sigma}$ and will correspond to monopoles and
antimonopoles (or vice versa, depending on orientation) in the base
$B$, which start life on the nucleation surface $B_{\Sigma}=\Sigma/U(1)$. The curves themselves are the Euclidean worldlines of the
monopoles. 

\begin{figure}[t]
\vspace{2pc}
\includegraphics[-30,0][100,130]{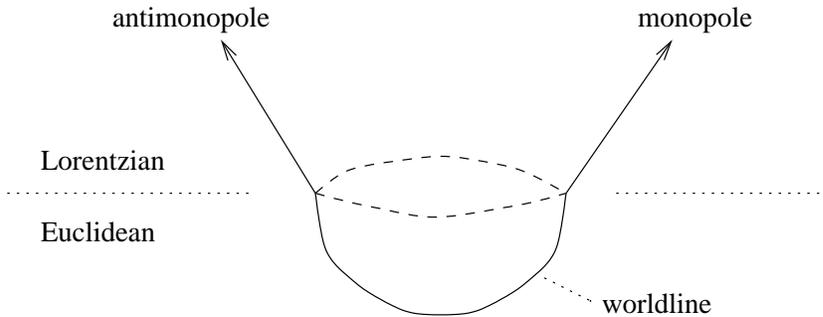}
\caption{Schematic picture of monopole-antimonopole creation. The dashed line lifts to the
nucleation surface $\Sigma$, and the solid line is the
monopole-antimonopole worldline, which lifts to a geodesic curve in $M\cup_{\Sigma}M_L$.}
\end{figure}

Note that this picture is qualitatively similar to electron-positron
pair production in a uniform electric field. In the Euclidean sector, one has
a single electron in flat space that travels round in a circle. One
may slice this circle along the equator and join onto the
post-tunnelling Lorentzian solution, in which one has an
accelerating electron-positron pair, pulled apart by the electric field. The electron and positron
are then viewed as the \emph{same} particle, the positron being viewed
as a
negatively charged electron travelling backwards in time, tunnelling
through the Euclidean sector, and re-emerging as an electron now
travelling forward in time. The Kaluza-Klein monopole picture above is very
similar, only here the worldline is identified with a stationary
geodesic $\gamma$ in $M\cup_{\Sigma}M_L$ and the charge is determined by the
linking number $\mathrm{link}(\gamma,\Sigma)$ of this
curve with the nucleation surface\footnote{The linking number is
defined to be $+1$ or $-1$ depending on whether $T_pM$ has the direct
sum orientation of $T_p\gamma\oplus{T}_p\Sigma$ or
not, where $p=\gamma\cap\Sigma$ \cite{pollack}.}.

Note that from this analysis we see that the number of
monopoles will always be equal to the number of antimonopoles; we will be
able to prove this later using the $G$-signature theorem, which refers
only to data on $\Sigma$. Note that one may also have closed geodesic curves in
$M$. These presumably correspond to virtual monopole-antimonopole
loops.

\subsection{The G-index theorem}

In this section we shall use the $G$-index theorem \cite{atiyah} to show
that the number of monopoles is always equal to the number of
antimonopoles (charge conservation), and that the total number of
defects produced is given by the Euler number of $\Sigma$. Note that
these statements are consistent with the fact that
$\chi(\Sigma)\in2\mathbb{Z}$. 

The $G$-index theorem is essentially a generalisation of the Lefschetz
fixed point theorem and the usual index theorem. Let $\mathcal{E}$ be
an elliptic complex on the compact manifold $\Sigma$, and suppose that the (topologically
cyclic\footnote{$G$ is topologically cyclic iff it contains an element
$g\in{G}$ such that the powers of $g$ are dense in $G$.}) compact Lie group $G$ acts on $\mathcal{E}$. Then
recall that the Lefschetz number $L(g,\mathcal{E})$ for a generator $g\in{G}$ of $G$ is defined as

\begin{equation}L(g,\mathcal{E})=\sum(-1)^i\mathrm{Tr}(g\mid{H}^i(\mathcal{E}))\label{lef}\end{equation}

where $H^i(\mathcal{E})$ are the homology groups of the complex
$\mathcal{E}$. The $G$-index theorem of \cite{atiyah} expresses the
Lefschetz number in terms of
the symbol of $\mathcal{E}$ and various characteristic classes,
evaluated over the fixed point set of $g$. The general formula is
rather complicated. The interested reader is referred to the original
paper \cite{atiyah}.

We will need to consider two 'classical' complexes; the de Rham complex,
and the signature complex. If $G$ acts on $\Sigma$, then it acts on
the latter complexes. We are of course interested in applying this to $G=U(1)$. Since $U(1)$
is connected, it acts trivially on the cohomology of $\Sigma$, and
hence in this case the Lefschetz number (\ref{lef}) reduces to the
usual index. The $G$-index theorem thus expresses the
Euler number and signature of $\Sigma$, being the indices of the de
Rham and signature complexes respectively, in terms of various
characteristic classes evaluated on the $U(1)$ stationary point
sets. This is actually reasonably straightforward. The details may be
found in the original papers \cite{atiyah}, but see also
\cite{eguchi}. We simply state the results. For the de Rham complex
one obtains

\begin{equation}\chi(\Sigma)=\sum_F\chi(F)\label{derham}\end{equation}

where the sum is over each connected component $F$ of the stationary point
set $\Sigma^{U(1)}$. 

The theorem for the signature complex is rather more
involved. However, it simplifies when the stationary points are
isolated. If $\Sigma$ has dimension $2r$ then one easily obtains

\begin{equation}\mathrm{Sign}(\Sigma)=\sum\prod_{j=1}^r(-i)\cot\left(\frac{n_j\tau}{2}\right)\label{sign}\end{equation}

The sum is over each isolated fixed point, with $\{n_j\}$
characterising the circle action on the normal bundle (which of course
is just the tangent bundle of $\Sigma$ restricted to the point) as in Section 3,
and the formula is valid for \emph{all} values of the group parameter
$\tau$. 

We now apply this to the case where $\Sigma$ has
$\mathrm{dim}(\Sigma)=4$ and trivial oriented cobordism class, so that $\mathrm{Sign}(\Sigma)=0$,
and the $U(1)$ action is semi-free, so that all $n_j$ are $\pm1$. We
choose our orientation convention by defining a monopole fixed point
 (or 'nut' in the terminology of \cite{gibhawk}) to have
$\prod_{j=1}^2n_j=+1$ and an antimonopole fixed point (or 'antinut') to have
$\prod_{j=1}^2n_j=-1$. The $G$-signature theorem thus states that the number of nuts minus the number
of antinuts is zero. Hence the number of monopoles is always equal to
the number of antimonopoles. This of course agrees with our earlier
discussion. There is, however, a slight subtlety in the argument. The
index theorem does not assume that $\Sigma$ bounds a manifold with
a smooth circle action, inducing the given circle action on
$\Sigma$. In principle such an extension might not exist. This
point is in any case irrelevant, since the existence of $M$ with
boundary $\Sigma$ is assumed from the outset. Nevertheless, it is interesting that the signature complex is related to
charge conservation in this way. 

The $G$-index theorem applied to the de Rham complex (\ref{derham})
gives us a less trivial result. Note that the Euler characteristic of
an oriented manifold is invariant under a change of
orientation. The Euler characteristic of a point is just one, and by
our previous remark is independent of how we choose to orient the point. Hence when all
stationary points are isolated, (\ref{derham}) implies that
$\chi(\Sigma)$ is equal to the total number of monopoles and
antimonopoles. This completes our analysis of the fixed point
theorems.

\subsection{Circle actions on $4$-manifolds and cobordism}

We now give a summary of a rather remarkable result due to
Fintuschel \cite{fint} on the classification of circle actions on
$4$-manifolds. We can then apply this classification to our nucleation
surface $\Sigma$, giving a complete description of the possible
topological configurations of monopole-antimonopole nucleation in
$D=5$ Kaluza-Klein theory. 

The main result of \cite{fint} is that if $\Sigma$ is a
simply-connected $4$-manifold admitting a smooth circle
action\footnote{In the original paper, the action is assumed to be
\emph{locally smooth.}. However, any smooth action of a compact Lie group
is locally smooth \cite{bredon}.}, then,
modulo the Poincar$\acute{\mathrm{e}}$ conjecture\footnote{Any 3-manifold that is
homotopy-equivalent to $S^3$ is homeomorphic to $S^3$. This remains as
probably one of the greatest unsolved problems in topology.}, $\Sigma$ is the connected sum of
copies of $S^4$, $S^2\times{S}^2$, $\mathbb{C}P^2$ and
$-\mathbb{C}P^2$. This rather deep result will be useful for
classifying, topologically, monopole-antimonopole production.

Recall that the connected sum $X\#Y$ of two closed oriented
$d$-manifolds $X$ and $Y$ is defined by removing (sufficiently
small) $d$-balls from each, and then identifying the boundaries of the
balls\footnote{One must of course smooth out the resulting space in
a neighbourhood of the identification. This can be done.}. It is easy to show that
$H_d(X\#Y;\mathbb{Z})\cong\mathbb{Z}$ and

\begin{equation}H_r(X\#Y;\mathbb{Z})\cong{H}_r(X;\mathbb{Z})\oplus{H}_r(Y;\mathbb{Z})\label{cohom}\end{equation}

for $0<r<d$.

It is clear that given a manifold $X$, $S^d\#X$ is
diffeomorphic to $X$. Thus the $S^4$ contribution in Fintuschel's Theorem is
rather trivial. 

We may now apply the theorem to our nucleation surface
$\Sigma$, assuming the latter is simply-connected. In many cases, one
might achieve this by going to a suitable covering space, although this
may result in a non-compact manifold. $T^4$ is the obvious example. Since $\Sigma$ is a boundary, its cobordism class is
trivial. Recall that two compact oriented $d$-manifolds $X$ and $Y$
belong to the same cobordism class \cite{stong} if and only if there is a compact oriented
$(d+1)$-manifold $M$ with boundary $\partial{M}$ such that $\partial{M}$, with its canonical orientation, is diffeomorphic to
$X+(-Y)$, where $+$ denotes the disjoint sum (distinguished from the
connected sum $\#$). One may then define an abelian group
$\Omega_d$ consisting of all oriented cobordism classes of
$d$-manifolds, the group composition being $+$. The low dimensional
oriented cobordism groups are conveniently listed in \cite{milnor}. In
particular, $\Omega_4\cong\mathbb{Z}$, the generator being
$\mathbb{C}P^2$. It follows that $\mathbb{C}P^2\#-\mathbb{C}P^2$ has
trivial cobordism class (that is, it bounds). In fact, a manifold $X$
is an oriented boundary if and only if all the Pontrjagin numbers\footnote{In
dimension four, one has $\mathrm{Sign}(X)=\frac{1}{3}p_1[X]$ where $p_1\in{H}^4(X;\mathbb{Z})$ is
the first Pontrjagin class of $X$.} and
Stiefel-Whitney numbers of $X$ vanish \cite{stong}. One may easily
verify that this is the case for
$\mathbb{C}P^2\#-\mathbb{C}P^2$. Hence one must have equal numbers of
$\mathbb{C}P^2$ and $-\mathbb{C}P^2$ in the classification theorem.

One may thus conclude that $\Sigma$ is either $S^4$ or the topological
sum of $S^2\times{S}^2$ and $\mathbb{C}P^2\#-\mathbb{C}P^2$. In fact,
there is an alternative way to view the latter. Recall that
$G$-bundles over $S^n$ are classified by $\pi_{n-1}(G)$
\cite{steenrod}. In the case $n=2$, we conclude that $S^2$ bundles
over $S^2$ are classified by $\pi_1(SO(3))\cong\mathbb{Z}_2$. Hence,
up to bundle isomorphism, there are two $S^2$ bundles over $S^2$; the
trivial bundle $S^2\times{S}^2$, and precisely one twisted
$S^2\times{S}^2$ product. The latter
is in fact diffeomorphic to $\mathbb{C}P^2\#-\mathbb{C}P^2$. This may
be seen via the explicit construction of the Page instanton
\cite{page} on the twisted $S^2$ bundle over $S^2$ in
\cite{pagepope}. Thus $\Sigma$ is $S^4$ or the connected sum of $S^2$
bundles over $S^2$.

If we impose that $\Sigma$ be a spin manifold, so that
the second Stiefel-Whitney class $w_2(\Sigma)\in{H}^2(\Sigma;\mathbb{Z}_2)$ vanishes, then neither
$\mathbb{C}P^2$ nor its orientation-reversed cousin $-\mathbb{C}P^2$
may contribute in the above. Thus, in this case, we have a 1-1
correspondence between the possible topologies of $\Sigma$ and the
Euler number of $\Sigma$. Specifically the possible Euler numbers for
$\Sigma$ are $2n$ ($n\in\mathbb{N}$) where $n=1$ corresponds to $S^4$ and $n>1$
corresponds to the connected sum of $n-1$ copies of
$S^2\times{S}^2$. This is easily computed from the formula
(\ref{cohom}). The Euler number is just the total number of isolated stationary points, fixed under some
semi-free circle action, which we may think
of as $n$ monopole-antimonopole pairs which start life on the
nucleation surface
$B_{\Sigma}=\Sigma/U(1)$. Note, however, that
each $\Sigma$ may admit many inequivalent circle actions, leading to different base spaces $B_{\Sigma}$.

\sect{Illustrations}

As an application of these results, we consider several explicit examples
describing the production of
Kaluza-Klein monopole-antimonopole pairs. Up until this point, we have largely been concerned with
topological issues. In the case that the nucleation surface is both
simply-connected and spin, we have seen that the possible topologies
are classified uniquely in terms of the total number of
monopoles. However, the possible physics described by this topological data is far from
unique. In order to produce pairs of objects, one needs some sort of physical
mechanism in order to pull the pair apart. For example, an electric
field in the case of electron-positron pairs, a positive cosmological
constant in the case of black holes, etc. As an illustration, we
consider two distinct physical mechanisms that support the production
of monopole-antimonopole pairs. In Section (6.1), we examine the case
of a positive cosmological constant. As is well known, this is a
source of gravitational repulsion, and is therefore able to support
the pair-production process. The topology of the nucleation surface is $S^4$,
and hence we produce a single pair of defects, in agreement with our
general analysis. In contrast, in Section (6.2) we consider
monopole-antimonopole production in the presence of a thin domain
wall. The
gravitational field produced by a domain wall is repulsive, and this
again causes the monopole-antimonopole pair to accelerate apart in the
Lorentzian sector. In Section (6.3) we consider a domain wall
configuration where the
topology of the nucleation surface is again $S^4$. Moreover, the solution is actually unique for
a given domain wall tension, if
we demand that the five-dimensional solution is flat either side of
the wall \cite{cham}. In Section (6.4) we construct a domain wall solution with nucleation
surface of topology $S^2\times{S}^2$. The total number of monopoles and
antimonopoles is $4$, again in agreement with our general
analysis.

\subsection{Cosmological constant}

Our starting point is the $(D+1)$-dimensional Euclidean gravitational
action\footnote{$M$ is closed, and note that we have changed our conventions
for the total dimension of spacetime (which is now Euclidean).}

\be
I_{\mathrm{EH}}=-\frac{1}{16\pi{G}_{D+1}}\int_M\mu(g_{D+1})\left(R(g_{D+1})-2\Lambda_{D+1}\right)\label{eh}\ee

where $R(g_{D+1})$ is the Ricci scalar of the $(D+1)$-metric $g_{D+1}$,$\mu(g_{D+1})$ denotes the
canonical Riemannian measure, $\Lambda_{D+1}>0$ is the cosmological
constant, and $G_{D+1}$ is the $(D+1)$-dimensional Newton
constant. Locally, one may choose coordinates on
$M$ adapted to the Killing vector field $k$ of the circle action $\Phi$. The metric $g_{D+1}$ then takes the
local form

\be
g_{(D+1),\mu\nu}dx^{\mu}dx^{\nu}=e^{\alpha\varphi}g_{D,ij}dx^idx^j+e^{\beta\varphi}(d\tau+C_1)^2\ee

where $0\leq\tau\leq2\pi{R}$ parametrises the circle of radius $R$, and all fields
are independent of $\tau$. The constants $\alpha$ and $\beta$ are
given by $\alpha=\sqrt{\frac{2}{(D-1)(D-2)}}$ and
$\beta=-(D-2)\alpha$. Note that if one tries to extend this local
coordinate system over a stationary point where $k=0$, the dilaton $\varphi$
necessarily diverges there. The action (\ref{eh}) reduces to the
\emph{Einstein} frame Einstein-Maxwell-dilaton form

\be
I_{\mathrm{EH}}=-\frac{1}{16\pi{G_D}}\int_B\mu(g_D)\left(R(g_D)-\frac{1}{2}(\partial\varphi)^2-\frac{1}{4}e^{-(D-1)\alpha\varphi}G_2^2-2\Lambda_{D+1}e^{\alpha\varphi}\right)\label{kk}\ee

where $G_2=dC_1$, $G_D=G_{D+1}/2\pi{R}$ and the base
$B^{\prime}=\left(M-M^{U(1)}\right)/U(1)$. Notice the potential term
$2\Lambda_{D+1}e^{\alpha\varphi}$.

As we have shown in Section 4, the
presence of Kaluza-Klein branes in the base $B$ requires that one adds
a Wess-Zumino coupling of the form (\ref{coupling}) to the \emph{string} frame
action, in order to reproduce the correct equation for the Kaluza-Klein
field strength, (\ref{brane}). Thus the
total effective action is given by

\be
I=I_{\mathrm{String}}+I_{\mathrm{WZ}}\ee

Of course, $I_{\mathrm{String}}=I_{\mathrm{Einstein}}\equiv{I}_{\mathrm{EH}}$
numerically; the string frame action is related to the Einstein frame
action by a field redefinition.

The maximally symmetric solution to the equations of motion obtained
from the action (\ref{eh}) is of course the $(D+1)$-dimensional
sphere, $S^{D+1}$, which may also be viewed as the Euclidean version
of de Sitter space, $dS_{D+1}$. The $(D+1)$-sphere metric

\be
ds^2_{D+1}=d\psi^2+\cos^2\psi{d}\Omega_D\ee

may therefore be regarded as an instanton for the creation of a
$(D+1)$-dimensional universe by a positive cosmological constant,
where, without loss of generality, we have set
$\Lambda=\frac{1}{2}D(D-1)$ in order to obtain a unit sphere as
solution. The
tunnelling manifold $M$ is given by $-\frac{\pi}{2}\leq\psi\leq0$
where $\psi=-\frac{\pi}{2}$ is the South pole of a
$(D+1)$-hemisphere, bounded by the equatorial $D$-sphere $\psi=0$,
which is zero-momentum and therefore may be taken as our nucleation
surface $\Sigma$. One may
obtain the Lorentzian section by analytically continuing $\psi=it$,
with $t>0$ giving the expanding de Sitter solution. 

We now specialise to the case $D=4$. Since $\Sigma=S^4$, it follows
that any smooth circle action will produce precisely one
monopole-antimonopole pair upon dimensional reduction. Specifically,
one may foliate the four-sphere by three-spheres, and take the Hopf
action on the latter

\be
d\Omega^2_4=d\rho^2+\frac{1}{4}\sin^2\rho\left((d\alpha+\cos\beta{d}\gamma)^2+d\beta^2+\sin^2\beta{d}\gamma^2\right)\ee

where $\rho$ is a polar coordinate on the four-sphere,
$0\leq\rho\leq\pi$, and $(\alpha,\beta,\gamma)$ are Euler angles on
the unit three-sphere. The circle action is generated by
$\partial/\partial\alpha$ which has isolated fixed points on the North
and South poles of the four-sphere, $\rho=0$, $\rho=\pi$,
respectively. These reduce to a monopole and antimonopole on the base
upon Kaluza-Klein reduction. The Euclidean worldline is identified with the
geodesic curve that starts at the North pole on the equatorial
four-sphere, moves back into $M$, intersecting the South pole of
the five-hemisphere at $\psi=-\frac{\pi}{2}$, and continuing back up
to the South pole on the equatorial four-sphere. In the Lorentzian
sector the monopole and antimonopole accelerate apart. This is illustrated
schematically in Figure 3.

\subsection{Domain walls}

We consider a thin domain wall $Y_{D+1}\subset{M}$, invariant under the circle action, with Nambo-Goto action

\be
I_{\mathrm{DW}}=\sigma_{D+1}\int_{Y_{D+1}}\mu(h_{D+1})+\frac{1}{8\pi{G}_{D+1}}\int_{Y_{D+1}}\mu(h_{D+1})[K_{D+1}]_+\label{dw}\ee

where $\sigma_{D+1}$ is the tension of the domain wall with
worldvolume $Y_{D+1}$, and $h_{D+1}$ is the induced metric on $Y_{D+1}$. Notice that $Y_{D+1}$ and the metric $h_{D+1}$, being a
hypersurface in the $(D+1)$-manifold $M$, actually have
dimension $D$, and not $D+1$. The label $D+1$ therefore refers to the
fact that objects live in the total space $M$, rather than denoting
their actual dimension. We hope this doesn't cause any confusion. We assume that the Killing vector field is
non-vanishing on $Y_{D+1}$. The unit normal $n_{D+1}$ to the domain wall points
\emph{into} $M$ on either side of the wall. The second
fundamental form of the imbedded hypersurface $Y_{D+1}$ is in general
discontinuous across the wall since the metric is continuous, but in
general non-differentiable, across $Y_{D+1}$. We must therefore
include a Gibbons-Hawking term in the action, summing the trace
$K_{D+1}$ of the
second fundamental form of $Y_{D+1}$ on either side of the wall. 

Upon dimensional reduction, the domain wall action (\ref{dw}) becomes

\be
I_{\mathrm{DW}}=\sigma_D\int_{Y_D}\mu(h_D)+\frac{1}{8\pi{G_D}}\int_{Y_D}\mu(h_D)[K_D+\frac{\alpha}{2}{n_D}.\partial\varphi]_+\ee

where $Y_D=Y_{D+1}/U(1)$ denotes the image of the worldvolume of the
invariant domain wall in the base, $h_D$ is the induced metric, $\sigma_D=2\pi{R}\sigma_{D+1}$ and $n_D$ is the unit normal (with respect
to $g_D$) of $Y_D$ in $B$, with trace of the second fundamental form
$K_D$.

We thus see that a domain wall of tension $\sigma_{D+1}$ in a purely gravitational $(D+1)$-dimensional
background may be viewed, upon Kaluza-Klein reduction, as a domain
wall of tension $\sigma_D$ in an Einstein-Maxwell-dilaton background.

Setting $\Lambda_{D+1}=0$, we see that varying the gravitational action

\be
I_{\mathrm{grav}}=I_{\mathrm{EH}}+I_{\mathrm{DW}}\label{action}\ee

with respect to the metric yields
the Israel matching conditions\footnote{For a nice derivation, see \cite{harvey}.}

\be
[K_{(D+1),\mu\nu}-K_{D+1}h_{(D+1),\mu\nu}]_+=8\pi{G_{D+1}}\sigma_{D+1}h_{(D+1),\mu\nu}\label{israel}\ee

One may find solutions to the equations derived from the action
(\ref{action}) as follows. One starts with a $(D+1)$-dimensional
Ricci-flat manifold $(M,g_{D+1})$ admitting a semi-free
isometric circle action. One then tries to find a totally umblic
invariant hypersurface
$Y_{D+1}\subset{M}$, that is
$K_{(D+1),\mu\nu}=ch_{(D+1),\mu\nu}$ for some constant $c$, that bounds some compact
region with boundary $Y_{D+1}$. One then constructs the
double $2M=M\cup_{Y_{D+1}}-M$. The hypersurface
$Y_{D+1}$ becomes a domain wall whose tension is determined by the
constant $c$. The Israel equations (\ref{israel}) are easily seen to
be satisfied provided

\be
c=-\frac{4\pi{G_{D+1}}}{D-1}\sigma_{D+1}=-\frac{4\pi{G_D}}{D-1}\sigma_D\ee

This provides us with a simple procedure for constructing domain wall
solutions.

Since $(2M,g_{D+1})$ is almost everywhere Ricci-flat, the
on-shell gravitational action becomes 

\be
I_{\mathrm{grav}}=-\frac{\sigma_{D+1}}{D-1}\mathrm{vol}(Y_{D+1})\label{onshell}\ee

\subsection{$\chi(\Sigma)=2$}

In this section, we analyse briefly the solution in
\cite{cham} and compute its action. Following the above procedure, one
starts with five-dimensional flat space $\mathbb{E}^5$

\be
ds_5^2=dr^2+r^2\left[d\psi^2+\frac{1}{4}\cos^2\psi\left((d\alpha+\cos\beta{d}\gamma)^2+d\beta^2+\sin^2\beta{d}\gamma^2\right)\right]\ee

The circle action is given by the Hopf action on the $S^3$ leaves that
foliate the $S^4$ principal orbits of the $SO(5)$ isometry group. That
is, we take the Killing vector field $\partial/\partial\alpha$. The
invariant umbilic hypersurface is given by $\{r=r_0>0\}$ and one easily
verifies that 

\be
r_0=\frac{3}{4\pi{G_5}\sigma_5}=\frac{3}{4\pi{G_4}\sigma_4}\ee

The gravitational action (\ref{onshell}) is then given by

\be
I_{\chi=2}=-\frac{2\pi}{3{G_5}}r_0^3\ee

The double $2M$ is almost everywhere flat and is topologically $S^5$.

The nucleation surface must be invariant and totally geodesic. One may
easily see that $\psi=0$ satisfies these requirements. The Euclidean
tunnelling manifold is then given by $-\frac{\pi}{2}\leq\psi\leq0$, the upper bound
corresponding to the nucleation surface $\Sigma\subset2M$, which is
topologically $S^4$. The circle action restricted to $\Sigma$
has two isolated fixed points, in agreement with our general
analysis. These are a nut and antinut, separated by the domain wall
restricted to the nucleation surface, and are located at
the two copies of $r=0$ either side of the wall. The worldline of the
monopole-antimonopole pair is the geodesic curve
$\psi=-\frac{\pi}{2}$. This runs from $r=0$ on the nucleation surface on one side of the wall,
intersects the domain wall at $r=r_0$, and then continues to the other
copy of $r=0$ on $\Sigma$.

One may obtain the Lorentzian solution by analytically continuing
$\psi=it$ with $t>0$. The monopole-antimonopole pair thus accelerate
apart in the Lorentzian section.

\subsection{$\chi(\Sigma)=4$}

In this section we construct a solution with two monopole-antimonopole
pairs. The nucleation surface, being spin, thus necessarily has the topology
$S^2\times{S}^2$. One starts with the five-dimensional Schwarzschild solution

\be
ds_5^2=\left(1-\left(\frac{r_H}{r}\right)^2\right)d\tau^2+\left(1-\left(\frac{r_H}{r}\right)^2\right)^{-1}dr^2+r^2\left[d\psi^2+\cos^2\psi(d\theta^2+\sin^2\theta{d}\phi^2)\right]\label{sch}\ee

Since the coordinate $\tau$ is identified with period $2\pi{r}_H$, we
see that the topology is $\mathbb{R}^2\times{S}^3$. We take the circle action generated by the Killing vector field
$k=\frac{\partial}{\partial\tau}+\frac{1}{r_H}\frac{\partial}{\partial\phi}$.
One may easily verify that there exists an invariant umbilic
hypersurface $\{r=r_0>0\}$ given by $r_0=\sqrt{2}r_H$, where the
tension $\sigma_5$ of the resulting domain wall is related to $r_0$ via

\be
\frac{r_0}{\sqrt{2}}=r_H=\frac{3}{8\pi{G}_5\sigma_5}=\frac{3}{8\pi{G}_4\sigma_4}\ee

The double $2M$ has topology
$S^2\times{S}^3$. Note that since the solution (\ref{sch}) is
regular only if $\tau\sim\tau+2\pi{r}_H$, we see that, in this case,
the radius of the Kaluza-Klein circle direction is related to the
tension of the domain wall. This is in contrast to the previous
example, where these quantities were independent. The gravitational action
(\ref{onshell}) is given by

\be
I_{\chi=4}=-\frac{\pi^2}{\sqrt{2}G_5}r_0^3\ee

The nucleation surface is given by $\psi=0$, and is topologically
$S^2\times{S}^2$. The tunnelling manifold is given by
$-\frac{\pi}{2}\leq\psi\leq0$, the upper bound corresponding to
$\Sigma$. The circle action generated by $k\mid_{\Sigma}$
has four isolated fixed points at the two copies of the two points
$\{r=r_H, \theta=0\}$ and $\{r=r_H, \theta=\pi\}$. One thus has two
monopole-antimonopole pairs separated by the domain wall restricted to
$\Sigma$. The mirror image of one monopole-antimonopole pair is
an antimonopole-monopole pair. The analytic continuation is again
given by setting $\psi=it$.

\sect{Conclusions}

The aim of this paper was to make more precise the relationship
between Kaluza-Klein branes and stationary points of circle
actions. In references \cite{horowitz} and \cite{refs}, various branes were
constructed as stationary point sets of circle actions, but the
relationship to the general theory of branes, and in particular the
cohomology equation (\ref{brane}), was unclear. Moreover, only
simple examples were presented. This motivated the
work in the present paper. In particular, in Section 4 we have shown quite
generally how codimension four stationary point sets may be
interpreted as magnetically charged branes in the reduced space, and, using an explicit
construction of the Thom class of the normal bundle of the brane in
the base, were able to prove the corresponding equation in cohomology (\ref{brane}). This puts the results of papers such as \cite{horowitz} in
a more general setting. Note, however, that since the dilaton diverges
as one approaches the brane, physically the spacetime decompactifies
in a neighbourhood of the brane worldvolume. So, strictly speaking,
the physics near the brane is goverened by the higher dimensional
theory. However, if one wishes to interpret the brane purely from the
lower-dimensional point of view, in order to make contact with the
general theory of branes, one must resort to a construction similar to
the one outlined in this paper.

In Section 5 we then went on to study the specific case of monopole-antimonopole production in a five-dimensional Kaluza-Klein
theory. Charge conservation and the number of defects produced are
related to various $G$-index theorems, and using Fintuschel's
classification of circle actions on simply-connected four-manifolds,
together with some simple cobordism results, we were able to classify
completely the possible topologies of the nucleation surface. Finally,
in Section 6 we gave several explicit examples of monopole-antimonopole nucleation,
where the production mechanism is supported either by a positive
cosmological constant, or a domain wall.

In conclusion, the theory of Kaluza-Klein branes is related to many
different areas of algebraic and differential topology; $G$-index
theorems, cobordism, and various characteristic classes all play an
important role in the general theory. This, together with recent work on the
relation of $K$-Theory to the physics of $D$-branes, suggests that perhaps there should exist a deeper and more fundamental mathematical
structure that underlies all of these ideas.

\medskip

\centerline{\bf Acknowledgments}

I am extremely grateful to Gary Gibbons for valuable discussions and comments,
and would like also to thank Burt Totaro and Stephen Hawking for useful
conversations, and Harvey Reall for comments on a preliminary draft.


\noindent


\begin{thebibliography}{99}

\bibitem{horowitz} F. Dowker, J.P. Gauntlett, G.W. Gibbons, G.T. Horowitz,
\emph{Nucleation of p-branes and fundamental strings}, Phys. Rev. D53
(1996), 7115-7128, and references therein.

\bibitem{refs}M.S. Costa, M. Gutperle, \emph{The Kaluza-Klein Melvin
solution in $M$-Theory}, JHEP 0103 (2001) 027

P.M. Saffin, \emph{Gravitating fluxbranes}, gr-qc/0104014

M. Gutperle, A. Strominger, \emph{Fluxbranes in string theory},
hep-th/0104136

M.S. Costa, C.A.R. Herdeiro, L. Cornalba,
\emph{Flux-branes and the dielectric effect in string theory},
hep-th/0105023,

R. Emparan, \emph{Tubular branes in fluxbranes},
hep-th/0105062,

\bibitem{gibbons}G.W. Gibbons, K. Maeda, \emph{Black holes and
membranes in higher dimensional theories with dilaton fields}, Nucl. Phys. B298
(1988) 741

\bibitem{cham}R. Caldwell, A. Chamblin, G.W. Gibbons, \emph{Pair
creation of black holes by domain walls}, Phys. Rev. D53 (1996) 7103-7114

\bibitem{hunter}S.W. Hawking, C.J. Hunter, \emph{Gravitational entropy
and global structure}, Phys. Rev. D59 (1999) 044025

\bibitem{fint}R. Fintushel, \emph{Circle actions on simply connected
$4$-manifolds}, Trans. Amer. Math. Soc. 230 (1977), 147-171 and \emph{Classification of circle actions on
$4$-manifolds}, Trans. Amer. Math. Soc. 242 (1978), 377-390

\bibitem{zippin}D. Montgomery, L. Zippin, \emph{Topological
transformation groups}, Wiley (Interscience), 1955

\bibitem{buc}For a recent proof, see B. Bucicovschi, \emph{Seeley's
Theory of Pseudodifferential Operators on Orbifolds}, math.DG/9912228

\bibitem{candelas}P. Candelas, M.Lynker, R. Schimmrigk,
\emph{Calabi-Yau manifolds in weighted} $\mathbb{P}^*_4$,
Nucl. Phys. B341 (1990), 383-402. See also D. Joyce \emph{A new construction of
compact 8-manifolds with holonomy Spin(7)}, J. Diff. Geom. 53 (1999), 89-130

\bibitem{atiyah}M.F. Atiyah, I.M. Singer, G.B. Segal, \emph{The index
of elliptic operators: I-III}, Ann. of Math. 87 (1968), 484-604

\bibitem{milnor}J.W. Milnor, J.D. Stasheff, \emph{Characteristic
classes}, Annals of Mathematics Studies No. 76, Princeton University Press

\bibitem{maunder}C.R.F. Maunder, \emph{Algebraic topology}, Cambridge
University Press 1980

\bibitem{pollack}V. Guillemin, A. Pollack, \emph{Differential
topology}, Prentice-Hall 1974

\bibitem{eguchi}T. Eguchi, P.B. Gilkey, A.J. Hanson, Phys. Rep. 66
(1980), 213

\bibitem{gibhawk}G.W. Gibbons, S.W. Hawking, \emph{Classification of
gravitational instanton symmetries}, Commun. Math. Phys. 66 (1979), 291-310

\bibitem{bredon}G.E. Bredon, \emph{Introduction to compact
transformation groups}, Academic Press, 1972

\bibitem{stong} R.E. Stong, \emph{Notes on cobordism theory},
Princeton University Press, 1968

\bibitem{steenrod} N. Steenrod, \emph{The topology of fibre bundles},
Princeton University Press, 1951

\bibitem{page} D.N. Page, \emph{A compact rotating gravitational
instanton}, Phys. Lett. 79B, No.3 (1978) 235-238

\bibitem{pagepope} D.N. Page, C.N. Pope, \emph{Inhomogenous Einstein
metrics on complex line bundles}, Class. Quantum Grav. 4 (1987) 213-225

\bibitem{hartle} G.W. Gibbons, J.B. Hartle, \emph{Real tunneling
geometries and the large-scale topology of the universe},
Phys. Rev. D42, No.8 (1990), 2458-2468

\bibitem{harvey}A.C. Chamblin, H.S. Reall, \emph{Dynamic dilatonic
domain walls}, Nucl. Phys. B562 (1999), 133-157

\bibitem{polchinski}J. Polchinski, \emph{Dirichlet-branes and
Ramond-Ramond charges}, Phys. Rev. Lett. 75 (1995) 4724-4727

\bibitem{bott}R. Bott, L.W. Tu, \emph{Differential forms in algebraic
topology}. Springer-Verlag 1982

\bibitem{witten1}E. Witten, \emph{Topological tools in 10-dimensional
physics}, Int. Journ. Mod. Phys. A, Vol.1 No.1 (1986), 39-64

\bibitem{witten2}G. Moore, E. Witten, \emph{Self-duality,
Ramond-Ramond fields, and K-Theory}, JHEP 0005 (2000) 032

\bibitem{ktheory}E. Witten, \emph{D-branes and K-Theory}, JHEP 9812
(1998), 019

\end{thebibliography}
\end{document}